\documentclass[onecolumn]{IEEEtran}
\renewcommand{\baselinestretch}{1.01}

\usepackage{graphicx}  % Written by David Carlisle and Sebastian Rahtz

\usepackage{amsmath}   % From the American Mathematical Society
\usepackage{amsfonts}

\usepackage{latexsym}

\newcommand{\mean}[1]{\langle{#1}\rangle}

\newcommand{\dgg}{^{\dagger}}
\newcommand{\Tr}{{\rm Tr}\hspace{0.07cm}}

\newcommand{\im}{{\rm i}}

\newcommand{\half}{\frac{1}{2}}

\newcommand{\condbar}{\hspace{0.1em}|\hspace{0.1em}}
\newcommand{\condbigbar}{\hspace{0.1em}\Big|\hspace{0.1em}}

\newtheorem{theorem}{Theorem}[section]
\newtheorem{lemma}{Lemma}[section]
\newtheorem{definition}{Definition}[section]
\newtheorem{example}{Example}[section]
\newtheorem{remark}{Remark}[section]

\begin{document}

\title{Quantum risk-sensitive estimation and robustness}


%%%%%%%%%%%%%%%%%%%%%%%%%%%%%%%%%%%%%%%%%%%%%%%%%%%%%%%%%%%%%%%%
%%%%%%%%%%%% Authors name and affiliation %%%%%%%%%%%%%%%%%%%%%%
%%%%%%%%%%%%%%%%%%%%%%%%%%%%%%%%%%%%%%%%%%%%%%%%%%%%%%%%%%%%%%%%

\author{Naoki Yamamoto$\mbox{}\dgg$ and Luc Bouten$\mbox{}^*$
% <-this % stops a spac
\thanks{
%* Supported by the ARO under Grant No. W911NF-06-1-0378. 
%}
$\dagger$
Department of Engineering, Australian National University, 
Canberra, ACT 2600, Australia (E-mail: {\tt naoki.yamamoto@anu.edu.au}). 
*Physical Measurement and Control MC:266-33, 
California Institute of Technology, Pasadena, CA 91125, USA 
(E-mail: {\tt bouten@its.caltech.edu}). 
}% <-this % stops a space
}
\maketitle

%~~~~~~~$\dagger$ 
%Address: Department of Engineering, Building 32, North Road, 
%Australian National University, 
%
%~~~~~~~\hspace{1.65cm}Acton ACT 0200, Australia 
%       ~~(correspondence address)
%
%~~~~~~~\hspace{0.29cm}Phone: +61-2-6125-3444, 
%       \hspace{0.29cm}Fax: +61-2-6125-0506
%
%~~~~~~~\hspace{0.29cm}E-mail: {\tt naoki.yamamoto@anu.edu.au}
%\\
%
%~~~~~~~* 
%Address: Physical Measurement and Control MC:266-33, 
%California Institute of Technology, 
%
%~~~~~~~\hspace{1.65cm}Pasadena, CA 91125, USA 
%
%~~~~~~~\hspace{0.29cm}Phone: +1-626-395-8431, 
%       \hspace{0.29cm}Fax: +1-626-395-3802 
%
%~~~~~~~\hspace{0.29cm}E-mail: {\tt bouten@its.caltech.edu}
%\\
%\\

%%%%%%%%%%%%%%%%%%%%%%%%%%%%%%%%%%%%%%%%%%%%%%%%%%%%%%%%%%%%%%%%%%%%%%%%
%%%%%%%%%%%%%%%%%%%%%%%%%%%%%%%%  Abstruct  %%%%%%%%%%%%%%%%%%%%%%%%%%%%%
%%%%%%%%%%%%%%%%%%%%%%%%%%%%%%%%%%%%%%%%%%%%%%%%%%%%%%%%%%%%%%%%%%%%%%%%%

\begin{abstract}

This paper studies a quantum risk-sensitive estimation problem and 
investigates robustness properties of the filter. 
This is a direct extension to the quantum case of analogous classical 
results. 
All investigations are based on a discrete approximation model of the 
quantum system under consideration. 
This allows us to study the problem in a simple mathematical setting. 
We close the paper with some examples that 
demonstrate the robustness 
of the risk-sensitive estimator. 

\end{abstract}

%\begin{keywords}
%IEEEtran, journal, \LaTeX, paper, template.
%\end{keywords}

%\begin{center} \bfseries EDICS Category: 3-BBND \end{center}
% Will be ignored for other modes.

\IEEEpeerreviewmaketitle

\section{Introduction}

Filtering, which in a broad sense 
is a method for extracting information 
from a noisy signal, is one 
of the principal tools in modern 
engineering science. 
In particular, when considering a partially observed dynamical system, 
we can construct an optimal filter that computes the least square estimate 
of a state variable of the dynamics. 
In the linear case, this results in 
the so-called {\it Kalman filter} \cite{kalman}. 
This dynamical filtering theory was rigorously established using the 
classical Kolmogorov probability theory and its application to the 
theory of stochastic differential equations (e.g.\ \cite{kushner}). 
Moreover, it is well known as the {\it separation theorem} \cite{wonham} 
that the solution of a general optimal control problem for a partially 
observed system can be represented in terms of a corresponding 
information state of the filter. 
For this reason, the filtering theory is not only important in itself, but 
also essential in feedback control theory.

The situation is much the same 
in quantum mechanics. 
The {\it Heisenberg uncertainty principle} 
shows that any quantum system must possess 
fundamental uncertainty originating from the 
noncommutativity of its random variables. 
Therefore, we can never have complete observation 
in the quantum setting, which implies the necessity of filtering 
in the quantum case. 
Fortunately, there exists a {\it quantum filtering theory} 
as a beautiful parallel to the classical one. 
The theory was pioneered by Belavkin in the remarkable papers 
\cite{belavkin1,belavkin2,belavkin3}, and 
the {\it quantum filtering equation} or {\it stochastic master equation} 
is now widely used in the physics community 
\cite{ahn,luc1,doherty,ramon1,mirrahimi,stockton,thomsen,naoki2}. 
Moreover, as in the classical theory, it is possible 
to show that a separation principle holds in the 
quantum case \cite{luc3}.

The filtering for both classical and quantum cases is, 
as mentioned above, clearly an important tool in control theory. 
However, we have to point out that the optimal filter is 
in general quite fragile 
to unmodeled uncertainty of the system, and consequently the optimal 
estimation can be largely violated. 
This fact requires us to develop a theory of {\it robust estimation} 
that allows some model uncertainties and guarantees high-quality 
estimation performance. 
{\it Guaranteed-cost filtering} \cite{petersen,xie} 
is one such robust estimation method in the classical theory. 
It guarantees that the variance of the estimation error is within a certain 
bound even when the linear system under consideration 
includes unknown parameters. 
Moreover, {\it risk-sensitive filtering} \cite{chara,dey,moore,speyer} 
is known as a very efficient robust estimation method, 
for a wide class of classical linear and nonlinear systems \cite{boel,
ugrinovskii,yoon}.  
Recently, one of the authors has obtained a 
quantum version of the guaranteed-cost filter 
mentioned above \cite{naoki1}. In this paper, 
we develop a quantum risk-sensitive estimation theory.

Let us first briefly introduce the classical theory of 
risk-sensitive estimation.

%%%%%%%%%%%%%%%%%%%%%%%%%%%%%%%%%%%%%%%%%%%%%%%%%%%%%%%%%%%%%%%%%%%
%%%%%%%%%%%%%%%%%%%%%%%%%%%%%%%%%%%%%%%%%%%%%%%%%%%%%%%%%%%%%%%%%%%

\subsection{Classical risk-sensitive estimation}

We are given a probability space $(\Omega,{\cal F},{\bf P})$ and 
a signal model of a discrete time system 
\begin{equation}
\label{classical-system}
    x_l=a(x_{l-1}) + b(x_{l-1}) w_l,~~~
    y_l=c(x_{l-1}) + v_l, 
\end{equation}
where $x_l$ is the signal state, $y_l$ is the output, and 
$w_l, v_l$ are {\it i.i.d.} random Gaussian processes. 
A version of the risk-sensitive estimator of $x_l$ is defined as 
\begin{equation}
\label{classical-rs-estimator}
    \hat{x}_l^\mu
      := \mathop{\mathrm{argmin}}_{z_l\in{\cal Y}_l}
             {\bf E}_{{\bf P}}[\Psi(z_l)],\qquad
    \Psi(z_l)={\rm exp}\Big[\mu_1\sum_{i=1}^{l-1}(x_i-\hat{x}_i^\mu)^2
                           +\mu_2(x_l-z_l)^2 \Big], 
\end{equation}
where ${\cal Y}_l=\sigma\{y_i ; 1\leq i\leq l\}$ is 
the $\sigma$-algebra generated from the observation 
$y_i$, and $\mu=(\mu_1, \mu_2)$ are the weighting 
constants called the {\it risk-sensitive parameters}. 
Moreover, we use the notation $z_l\in{\cal Y}_l$ 
to indicate that $z_l$ is a bounded ${\cal Y}_l$-measurable 
function. 
The risk-sensitive estimator (\ref{classical-rs-estimator}) can be 
represented by 
$\hat{x}_l^\mu
=\mathop{\mathrm{argmin}}_{z_l\in{\cal Y}_l}f_1(z_l, \alpha_l^\mu)$, 
where $f_1$ is a certain function and $\alpha_l^\mu(x)$ is an 
{\it information state} defined by 
\begin{equation}
\label{classical-info-state}
    {\bf E}_{{\bf Q}}\Big[
       \Lambda_l \zeta(x_l)
          {\rm e}^{\mu_1\Sigma_{i=1}^{l-1} (x_i-\hat{x}_i^\mu)^2}
             \condbigbar{\cal Y}_l\Big]
    =\int_{{\mathbb R}}\zeta(x)\alpha_l^\mu(x)dx, 
\end{equation}
for all test functions $\zeta(x)$. 
Here, ${\bf Q}$ is a probability measure defined by 
\begin{equation}
\label{classical-change-of-measure}
    \Lambda_l
    :=\frac{d{\bf P}}{d{\bf Q}}
    =\prod_{i=0}^l
    {\rm exp}\Big[c(x_i)y_i-\frac{1}{2}c(x_i)^2\Big]. 
\end{equation}
Moreover, $\alpha_l^\mu(x)$ satisfies a recursive equation of the form 
$\alpha_l^\mu=f_2(\alpha_{l-1}^\mu,\hat{x}_{l-1}^\mu,y_l)$. 
Hence, running this equation with the measurement data $y_l$, we can 
recursively calculate $f_1(z_l,\alpha_l^\mu)$ and obtain the minimizer of 
this function, i.e., $\hat{x}_l^\mu$.

Note that $\hat{x}_l^\mu$ differs from the standard optimal 
(or {\it risk-neutral}) estimator 
$\hat{x}_l:=\mathop{\mathrm{argmin}}_{z_l\in{\cal Y}_l}$
${\bf E}_{{\bf P}}[(x_l-z_l)^2]$ and is thus not optimal in the sense of 
the mean square error. 
However, the risk-sensitive estimator certainly has a great advantage 
over the risk-neutral one when we consider an uncertain system. 
This can be seen as follows. 
If the true probability measure ${\bf P}_{{\rm true}}$ 
is unknown, then we need to use a known nominal 
measure ${\bf P}_{{\rm nom}}$ and design 
a nominal filter based on ${\bf P}_{{\rm nom}}$. 
However, since ${\bf P}_{{\rm true}}\neq{\bf P}_{{\rm nom}}$, 
there is no guarantee that the nominal estimator 
$\hat{x}_l^{{\rm nom}}$ yields 
a bounded estimation error. 
The risk-sensitive estimator overcomes this issue. 
That is, the nominal risk-sensitive estimator 
$\hat{x}_l^{\mu,{\rm nom}}$ (i.e., based on 
${\bf P}_{{\rm nom}}$) satisfies 
\begin{eqnarray}
\label{classical-robustness}
& & \hspace*{-1em}
    {\bf E}_{{\bf P}_{{\rm true}}}\big[ 
       \mu_1\sum_{i=1}^{l-1}(x_i-\hat{x}_i^{\mu,{\rm nom}})^2
       +\mu_2(x_l-\hat{x}_l^{\mu,{\rm nom}})^2 \big]
\nonumber \\ & & \hspace*{3em}
    \leq 
    \log{\bf E}_{{\bf P}_{{\rm nom}}}
                           [\Psi(\hat{x}_l^{\mu,{\rm nom}})]
       +R^c({\bf P}_{{\rm true}}\|{\bf P}_{{\rm nom}}), 
\end{eqnarray}
where $R^c({\bf Q}\|{\bf P}):=\int\log(d{\bf Q}/d{\bf P})d{\bf Q}$ is the 
classical relative entropy of ${\bf Q}$ and ${\bf P}$. 
Eq.\ \eqref{classical-robustness} implies that the unknown 
true estimation error is bounded if 
$R^c({\bf P}_{{\rm true}}\|{\bf P}_{{\rm nom}})$ is finite. 
This robustness property is derived using the following {\it duality 
relation} (e.g.\ \cite{dupuis}) of two measures ${\bf P}$ and ${\bf Q}$: 
\begin{equation}
\label{classical-duality}
    \log{\bf E}_{{\bf P}}[{\rm e}^\psi]
     =\sup_{{\bf Q}}\Big[
         {\bf E}_{{\bf Q}}(\psi)-R^c({\bf Q}\|{\bf P})
            ~:~{\bf Q}\ll{\bf P}\Big], 
\end{equation}
where ${\bf Q}\ll{\bf P}$ means that ${\bf Q}$ is absolutely continuous to 
${\bf P}$.

%%%%%%%%%%%%%%%%%%%%%%%%%%%%%%%%%%%%%%%%%%%%%%%%%%%%%%%%%%%%%%%%%%%
%%%%%%%%%%%%%%%%%%%%%%%%%%%%%%%%%%%%%%%%%%%%%%%%%%%%%%%%%%%%%%%%%%%

\subsection{Organization of the paper}

This paper provides a quantum version of the risk-sensitive 
estimation method presented above and shows its robustness 
properties against system uncertainty. The systems 
we consider are taken from quantum optics and consist 
of a quantum system in interaction with the quantized 
electromagnetic field. The field is described by a 
discretized model \cite{luc2} that converges to a quantum 
stochastic dynamics \cite{hudson} when the discretization 
step is taken to zero \cite{attal1,attal2,brun,gough,lindsay}.
The discretized model has the advantage of being 
very tractable mathematically.
The estimator is based on the risk-sensitive 
information state introduced by James \cite{dhelon,james1} 
in the context of quantum risk-sensitive control. 
We derive a bound on the estimation error in the 
presence of uncertainty.
We illustrate the robustness of the estimator by 
simulations.

The paper is organized as follows. 
In Section II we introduce quantum 
probability in a finite dimensional 
context and a duality relation that 
will lead to the robustness property of 
the estimator. 
Section III is devoted to describe 
a discrete approximation model 
of the field. Section IV introduces 
the notion of composition of an 
operator  and an operator valued function.
In Section V we introduce the risk-sensitive 
estimator and derive the filter propagating 
the risk-sensitive information state. 
Section VI introduces a class of uncertain 
systems and derives a bound on the estimation 
error, showing robustness. In Section VII 
we present the results from simulations.

%%%%%%%%%%%%%%%%%%%%%%%%%%%%%%%%%%%%%%%%%%%%%%%%%%%%%%%%%%%%%%%%%%%
%%%%%%%%%%%%%%%%%%%%%%%%%%%% Q probability %%%%%%%%%%%%%%%%%%%%%%%%
%%%%%%%%%%%%%%%%%%%%%%%%%%%%%%%%%%%%%%%%%%%%%%%%%%%%%%%%%%%%%%%%%%%

\section{Quantum probability theory}

\subsection{Quantum probability space}

In quantum mechanics, a random variable is represented by a linear 
self-adjoint operator on a Hilbert space. Due to the noncommutativity 
of such operators, we need to replace
the conventional notion of a classical probability 
space $(\Omega,{\cal F},{\bf P})$ by the notion of a  
{\it quantum probability space} defined below.

\begin{definition}[$*$-algebra] 
\label{*-algebra}
Let ${\mathsf H}$ be a finite-dimensional complex Hilbert space. 
A $*$-{\it algebra} ${\cal A}$ is a set of linear operators 
${\mathsf H}\rightarrow{\mathsf H}$ such that 
$I, \alpha A+\beta B, AB, A^*\in{\cal A}$ 
for any $A,B\in{\cal A}$ and $\alpha,\beta \in {\bf C}$.  
${\cal A}$ is called commutative if $[A,B]=AB-BA=0$ for any 
$A,B\in{\cal A}$. 
\end{definition}

\begin{definition}[State] 
\label{State}
A {\it state} on ${\cal A}$ is a linear map 
${\mathbb P}:{\cal A}\rightarrow{\bf C}$ that is positive 
${\mathbb P}(A^*A)\geq 0,~\forall A\in{\cal A}$ and normalized 
${\mathbb P}(I)=1$. 
\end{definition}

Let $d$ be the dimension of $\mathsf{H}$.
Let $(e_1, \ldots, e_d)$ be 
an orthonormal basis of $\mathsf{H}$. The {\it trace}  
is the state defined by 
$\mbox{Tr}(A) = \sum_{i=1}^d \langle e_i, A e_i\rangle$ 
for all $A \in \mathcal{A}$. It is well known that 
this definition does not depend on the basis.

\begin{definition}[Quantum probability space] 
\label{Quantum probability space}
Let ${\cal A}$ be a $*$-algebra of operators on a finite-dimensional 
complex Hilbert space ${\mathsf H}$ and ${\mathbb P}$ be a state on 
${\cal A}$. 
Then, $({\cal A},{\mathbb P})$ is called a (finite-dimensional) 
quantum probability space. 
\end{definition}

Let $({\cal A},{\mathbb P})$ be a quantum probability space. 
A self-adjoint element of ${\cal A}$ is called a 
{\it quantum random variable} or {\it observable}.
If ${\cal A}$ is a commutative $*$-algebra, 
then we call $({\cal A},{\mathbb P})$ 
a commutative quantum probability space. 
In this case, all quantum random variables in ${\cal A}$ commute with 
each other, which is the same as in the classical case. 
It is therefore not surprising that a commutative quantum probability 
space is equivalent to a classical one. 
A formal statement of this assertion is provided by 
the well known {\it spectral theorem} (Theorem \ref{Spectral theorem} below).
Note that in the finite dimensional setting of this 
article the spectral theorem follows trivially from 
diagonalizing the operators in ${\cal A}$ (see the 
proof of Theorem \ref{Spectral theorem} below). In an 
infinite dimensional setting an analogous result, 
which is closely related to Gelfand's 
Theorem for commutative $C^*$-algebras 
(see e.g.\ \cite{sakai}), is true.

\begin{definition}[$*$-isomorphism]
\label{*-isomorphism}
Let $\Omega$ be a set and let ${\cal F}$ be 
a $\sigma$-algebra on $\Omega$. 
A $*$-{\it isomorphism} between a commutative $*$-algebra ${\cal C}$ and the 
set of bounded ${\cal F}$-measurable functions 
$\ell^\infty({\cal F})$ on $\Omega$ is a linear 
bijection $\iota:{\cal C}\rightarrow \ell^\infty({\cal F})$ such that 
$\iota(A^*)(i)=\iota(A)(i)^*$ and 
$\iota(AB)(i)=\iota(A)(i)\iota(B)(i)$ 
for all $A,B\in{\cal C}$ and $i\in\Omega$. 
\end{definition}

\begin{theorem}[Spectral theorem] 
\label{Spectral theorem}
Let $({\cal C},{\mathbb P})$ be a finite-dimensional commutative quantum 
probability space. 
Then there exists a classical probability space $(\Omega,{\cal F},{\bf P})$ 
and a $*$-isomorphism $\iota:{\cal C}\rightarrow\ell^\infty({\cal F})$ such 
that ${\mathbb P}(A)={\bf E}_{{\bf P}}[\iota(A)],~\forall A\in{\cal C}$. 
\end{theorem}

\begin{proof}
The theorem is proved by construction. 
First, let ${\mathsf H}={\mathbb C}^n$ and $\Omega=\{1,\ldots,n\}$. 
Since $[A, A^*]=0~\forall A\in{\cal C}$, all the elements in ${\cal C}$ 
can be diagonalized simultaneously. 
Hence, we can set $A={\rm diag}\{a_1,\ldots,a_n\}$ and define a classical 
random variable $\iota(A): \Omega\rightarrow{\mathbb C}$ by 
$\iota(A)(i)=a_i$. Let $P$ be a projection in ${\cal C}$, i.e.,
$P=P^*=P^2$, then $\iota(P)$ is the indicator 
function of a subset $S_P$ of $\Omega$. We define 
${\cal F}$ as the set of subsets $S_P$ of $\Omega$ 
where $P$ runs through the projections in ${\cal C}$.
Furthermore, we define a probability measure ${\bf P}$ on 
${\cal F}$
by ${\bf P}(S_P)={\mathbb P}(P),~\forall P\in{\cal C}$. 
As a result, we have constructed a classical probability space 
$(\Omega,{\cal F},{\bf P})$. 
It is easy to verify 
${\bf E}_{\bf P}[\iota(A)]={\mathbb P}(A)$. 
\end{proof}

Note here that any observable $A=A^*\in{\cal A}$ 
is an element of the commutative $*$-subalgebra 
${\cal C}\subset{\cal A}$ generated by $A$ 
itself. Using the spectral theorem we see 
that we can always realize an observable $A$
as a classical random variable $\iota(A)$ 
on a classical probability space $(\Omega,{\cal F},{\bf P})$, 
where the measure ${\bf P}$ is given by the state. 
If we perform a {\it measurement} of $A$, 
we obtain one of the values that 
$\iota(A)$ 
can take, distributed according to ${\bf P}$. Note 
that if two observables do not commute with each 
other, then we cannot represent them both as 
classical random variables on the same probability 
space. Such observables are called {\it incompatible}, 
they cannot both be measured in a single 
realization of an experiment.

\begin{example}[Quantum two-level system] 
\label{Quantum two-level system}
Let ${\mathsf H}={\mathbb C}^2$ and 
let ${\cal M}$ be the $*$-algebra of $2\times 2$ complex matrices.
Moreover, let $\psi$ be a state on ${\cal M}$. 
With the quantum probability space 
$({\cal M}, \psi)$ we can model a
{\it two-level system}. 
The state $\psi$ can be written as 
$\psi(X) = \Tr(\rho A),~\forall A \in {\cal M}$ 
for some operator $\rho$ that is positive and normalized 
(i.e., $\Tr(\rho) = 1$). 
Let us now consider a commutative $*$-subalgebra 
${\cal D}=\{D={\rm diag}\{d_1,d_2\}\condbar d_1,d_2\in{\bf R}\}
\subset{\cal M}$. 
From Theorem \ref{Spectral theorem}, we can construct 
a classical probability space that is in 
one-to-one correspondence with $({\cal D},\psi)$. 
The sample space is $\Omega=\{1,2\}$, and the set of events is 
${\cal F}=\{ \emptyset, \{1\}, \{2\}, \Omega \}$. 
A classical random variable $\iota(D)$ is then defined through 
$\iota(D)(1)=d_1$ and $\iota(D)(2)=d_2$. 
Now, $D\in{\cal D}$ has a spectral decomposition $D=\sum d_i P_i$ with the 
projection matrices $P_1={\rm diag}\{1,0\}$ and $P_2={\rm diag}\{0,1\}$, 
which yield classical indicator functions $\chi_{\{1\}}=\iota(P_1)$ and 
$\chi_{\{2\}}=\iota(P_2)$. 
Hence, the probability distribution of $\iota(D)$ is given by 
${\bf P}(\{1\})=\psi(P_1)=\Tr(\rho P_1)=\rho_{11}$ and 
${\bf P}(\{2\})=\rho_{22}$. 
\end{example}

Let $({\cal A}_1, {\mathbb P}_1)$ and 
$({\cal A}_2, {\mathbb P}_2)$ be two quantum probability 
spaces, defined on the Hilbert spaces 
${\mathsf H}_1$ and ${\mathsf H}_2$, respectively.
We will now introduce the composite quantum probability 
space $({\cal A}_1\otimes{\cal A}_2, 
{\mathbb P}_1\otimes{\mathbb P}_2)$.
Let $a_1\otimes a_2$ be the tensor (Kronecker) product of two 
vectors $a_1\in{\mathsf H}_1$ and $a_2\in{\mathsf H}_2$. 
Introducing an inner product 
$\mean{a_1\otimes a_2, b_1\otimes b_2}:=\mean{a_1, b_1}\mean{a_2, b_2}$, 
we have a Hilbert space ${\mathsf H}_1\otimes{\mathsf H}_2$. 
The composite quantum probability space 
$({\cal A}_1\otimes{\cal A}_2, {\mathbb P}_1\otimes{\mathbb P}_2)$ is 
then defined on ${\mathsf H}_1\otimes{\mathsf H}_2$ as follows. 
First, we define an element $A_1\otimes A_2\in{\cal A}_1\otimes{\cal A}_2$ 
through the relation 
$(A_1\otimes A_2)(a_1\otimes a_2)=A_1 a_1\otimes A_2 a_2$. 
Any element of ${\cal A}_1\otimes{\cal A}_2$ is given as a linear combination 
of such elements. 
Second, the state ${\mathbb P}_1\otimes{\mathbb P}_2$ is defined by 
$({\mathbb P}_1\otimes {\mathbb P}_2)(A_1\otimes A_2)
={\mathbb P}_1(A_1){\mathbb P}_2(A_2)$.

%%%%%%%%%%%%%%%%%%%%%%%%%%%%%%%%%%%%%%%%%%%%%%%%%%%%%%%%%%%%%%%%%%%
%%%%%%%%%%%%%%%%%%%%%%%%%%%%%%%%%%%%%%%%%%%%%%%%%%%%%%%%%%%%%%%%%%%

\subsection{Conditional expectation}\label{sec condexp}

Let $({\cal A}, {\mathbb P})$ be a quantum 
probability space. Let $A$ and $B$ be two 
commuting self-adjoint elements of ${\cal A}$. Using Theorem 
\ref{Spectral theorem} we can present $A$ and $B$
as classical random variables $\iota(A)$ and 
$\iota(B)$ on a classical probability space 
$(\Omega,{\cal F},{\bf P})$. This allows us to form 
the classical conditional expectation 
${\bf E}[\iota(A)\condbar\iota(B)]$. The {\it quantum 
conditional expectation} ${\mathbb P}(A|B)$ 
can then be defined as its pull-back 
\begin{equation*}
    {\mathbb P}(A\condbar B)
    =\iota^{-1}\Big(
          {\bf E}_{{\bf P}}[\iota(A)\condbar\iota(B)]\Big). 
\end{equation*}
Now suppose that instead of the operator $B$, 
we want to condition $A$ on a commutative 
$*$-subalgebra ${\cal C}$ of ${\cal A}$. As long as 
$A$ commutes with every element in ${\cal C}$, we can 
apply the spectral theorem to the commutative 
$*$-algebra generated by ${\cal C}$ and $A$ together, 
and define 
\begin{equation}
\label{Qcond-expectation}
    {\mathbb P}(A\condbar{\cal C})
    =\iota^{-1}\Big(
          {\bf E}_{{\bf P}}[\iota(A)\condbar\sigma(\iota({\cal C}))]\Big), 
\end{equation}
where 
$\sigma(\iota({\cal C}))$ stands for the classical $\sigma$-algebra 
generated by $\iota(K),~K\in{\cal C}$. 
This shows that given a commutative $*$-subalgebra 
${\cal C}$, we can define the quantum 
conditional expectation onto ${\cal C}$ for 
every self-adjoint element in the {\it commutant} of 
${\cal C}$. Here the commutant of 
${\cal C}$ is given by
\[
    {\cal C}':=\{ A\in{\cal A}~|~[A,C]=0~\forall C\in{\cal C} \}. 
\]
The formal definition of the conditional 
expectation follows below. It coincides with 
the standard definition of the conditional 
expectation for operator algebras \cite{umegaki, tomiyama} 
for the situation we are interested in. 
Note, however, that our definition is more restrictive 
since we only allow for conditional expectations 
from the commutant of a commutative algebra ${\cal C}$ onto 
${\cal C}$ itself.

\begin{definition}[Quantum conditional expectation] 
\label{Quantum conditional expectation}
Let $({\cal A},{\mathbb P})$ be a quantum probability space, and let 
${\cal C}$ be a commutative $*$-subalgebra of ${\cal A}$. 
Then the map 
${\mathbb P}(\,\cdot\,\condbar{\cal C}):{\cal C}'\rightarrow{\cal C}$ is 
called (a version of) the quantum conditional expectation from ${\cal C}'$ 
to ${\cal C}$ if 
${\mathbb P}({\mathbb P}(A\condbar{\cal C})K)={\mathbb P}(AK)~
\forall A\in{\cal C}',~\forall K\in{\cal C}$. 
\end{definition}

Note that for every self-adjoint 
element $A\in{\cal C}'$, we have 
given an explicit expression 
for the quantum conditional expectation
in Eq.\ \eqref{Qcond-expectation}. Every 
element $A$ in the commutant can be written in a unique way 
as $A = A_1+iA_2$ with $A_1$ and $A_2$ self-adjoint.
If we define the conditional expectation of 
$A$ onto ${\cal C}$ by linear extension 
of the definition in Eq.\ \eqref{Qcond-expectation}, 
then it is easy to see that it satisfies 
the formal definition given in Definition 
\ref{Quantum conditional expectation}. 
This means we have shown existence of the 
quantum conditional expectation as defined 
in Definition \ref{Quantum conditional expectation}.

Finally, we remark some basic properties that 
both the classical and quantum 
conditional expectations satisfy: 
${\rm (i)}$ ${\mathbb P}(A\condbar{\cal C})$ is unique with probability one, 
${\rm (ii)}$ ${\mathbb P}({\mathbb P}(A\condbar{\cal C}))={\mathbb P}(A)$, 
${\rm (iii)}$ ${\mathbb P}(CA\condbar{\cal C})=C\mathbb{P}(A\condbar{\cal C})$ 
if $C\in{\cal C}$ and $A \in \cal{C}'$ (module property), and   
${\rm (iv)}$ ${\mathbb P}({\mathbb P}(A\condbar{\cal B})\condbar{\cal C}) 
={\mathbb P}(A\condbar{\cal C})$ if ${\cal C}\subset{\cal B}$ 
(tower property). 
Note that it easily follows from the tower property that 
$\mathbb{P}(\,\cdot\,\condbar {\cal C})$ is idempotent, i.e.\ it 
is a projection.
Moreover, similar to the classical case, ${\mathbb P}(A\condbar{\cal C})$ 
is the least mean square estimate of $A$ given ${\cal C}$, i.e., 
\begin{equation}
\label{Qcond-LSE}
    \|A-{\mathbb P}(A\condbar{\cal C})\|_{{\mathbb P}}
    \leq \|A-{\mathbb P}(A\condbar{\cal C})\|_{{\mathbb P}}
        +\|{\mathbb P}(A\condbar{\cal C})-B\|_{{\mathbb P}}
    = \|A-B\|_{{\mathbb P}}~~\forall B\in{\cal C}, 
\end{equation}
where we have defined $\|X\|_{{\mathbb P}}^2:={\mathbb P}(X^*X)$.

%%%%%%%%%%%%%%%%%%%%%%%%%%%%%%%%%%%%%%%%%%%%%%%%%%%%%%%%%%%%%%%%%%%
%%%%%%%%%%%%%%%%%%%%%%%%%%%%%%%%%%%%%%%%%%%%%%%%%%%%%%%%%%%%%%%%%%%

\subsection{Density matrix and quantum relative entropy}

In Example \ref{Quantum two-level system}, we have seen that 
the state ${\mathbb P}$ can be represented 
in terms of a matrix $\rho$. 
In the finite dimensional case  
we can always find a unique {\it density matrix} $\rho$ that satisfies 
\begin{equation}
\label{density-matrix}
    {\mathbb P}(A)=\Tr(\rho A),~~
    \rho=\rho^*\geq 0,~~
    \Tr\rho=1. 
\end{equation}
The latter two conditions guarantee 
${\mathbb P}(A^*A)\geq 0~\forall A\in{\cal A}$ and ${\mathbb P}(I)=1$, 
respectively. 
In particular, when $\rho$ is a rank-one 
projection matrix $\rho=bb^*, 
b\in{\mathsf H}$, then ${\mathbb P}(A)$ 
is expressed as 
\begin{equation}
\label{pure-state}
    {\mathbb P}(A)=\Tr(bb^* A)=\mean{b, Ab}, 
\end{equation}
where $\mean{\,\cdot\,,\,\cdot\,}$ denotes 
the standard Euclidean inner product of 
two vectors.

In analogy to the classical relative entropy, which 
has been introduced in Section I, we can define
the {\it quantum relative entropy} of two states
in terms of their density matrices as
\begin{equation}
\label{QrelativeEntropy}
    R(\rho\|\rho'):=\Tr\big[ \rho(\log\rho-\log\rho') \big]~~
    \mbox{if}~~{\rm supp}\rho\subseteq{\rm supp}\rho', 
\end{equation}
where ${\rm supp}\rho$ represents the 
linear space spanned by the eigenvectors of 
$\rho$ \cite{ohya}. 
If ${\rm supp}\rho\subseteq\hspace{-0.9em}\slash~{\rm supp}\rho'$, 
then $R(\rho\|\rho'):=+\infty$. 
A quantum version of the duality relation 
(\ref{classical-duality}) is given as follows.

\begin{lemma}[Duality, see e.g.\ \cite{ohya} Prop.~1.11] 
\label{duality}
For any observable $A\in{\cal A}$ and density matrices $\rho$ and $\rho'$, 
the following relation holds: 
\begin{equation}
\label{Q-duality}
    \log\Tr({\rm e}^{A+\log\rho'})
    =\max_{\rho}\Big[
            \Tr(\rho A)-R(\rho\|\rho')~:~
                 {\rm supp}\rho\subseteq{\rm supp}\rho' \Big]. 
\end{equation}
\end{lemma}

\begin{proof}
The proof is straightforward. 
Defining a density matrix 
$\rho_o={\rm e}^{A+\log\rho'}/\Tr[{\rm e}^{A+\log\rho'}]$, we obtain 
\[
    \Tr(\rho A)-R(\rho\|\rho')
     =\log\Tr({\rm e}^{A+\log\rho'})-R(\rho\|\rho_o). 
\]
Then, as the quantum relative entropy $R(\rho\|\rho_o)$ is always 
non-negative and takes zero only when $\rho=\rho_o$, we observe that 
Eq.\ \eqref{Q-duality} holds and the maximum is attained only when 
$\rho=\rho_o$. 
\end{proof}

We can derive a relaxed form of Eq.\ \eqref{Q-duality}, expressed 
in terms of the corresponding states. 
From the Golden-Thompson inequality 
$\Tr({\rm e}^{A+B})\leq\Tr({\rm e}^A{\rm e}^B)$ (see \cite{golden,thompson}) 
with $A,B$ self-adjoint, we have 
\[
    \log\Tr({\rm e}^{A+\log\rho'})
    \leq\log\Tr({\rm e}^{A}{\rm e}^{\log\rho'})
    =\log\Tr({\rm e}^{A}\rho'). 
\]
Therefore, denoting the states corresponding 
to $\rho$ and $\rho'$ by ${\mathbb P}$ and 
${\mathbb P}'$ respectively, we have 
\begin{equation}
\label{duality-ineq}
    {\mathbb P}(A) \leq \log{\mathbb P}'({\rm e}^A) + R(\rho\|\rho'). 
\end{equation}
This inequality will be used to show robustness properties of the 
quantum risk-sensitive estimator.

%%%%%%%%%%%%%%%%%%%%%%%%%%%%%%%%%%%%%%%%%%%%%%%%%%%%%%%%%%%%%%%%%%%
%%%%%%%%%%%%%%%%%%%%%%%%%%%% Q filtering %%%%%%%%%%%%%%%%%%%%%%%%%%
%%%%%%%%%%%%%%%%%%%%%%%%%%%%%%%%%%%%%%%%%%%%%%%%%%%%%%%%%%%%%%%%%%%

\section{The discrete field and quantum filtering}

In this paper we restrict ourselves to a 
system that consists of a two-level atom 
in interaction with the quantized electromagnetic 
field. This is merely for reasons of 
convenience, the theory can easily be 
extended to a large class of systems in 
interaction with the field. In this Section 
we first introduce a discrete model for the electromagnetic 
field (see \cite{luc2} and the references therein). 
Second, we describe the interaction 
between the atomic system and the field. Due 
to the interaction, the field carries off 
information about the system. In this way, 
by measuring the field, we can perform a 
noisy observation of the system. Finally, 
using quantum filtering theory we 
form optimal estimates of the atom observables. 
The quantum filtering equation recursively propagates these 
estimates in time.

%%%%%%%%%%%%%%%%%%%%%%%%%%%%%%%%%%%%%%%%%%%%%%%%%%%%%%%%%%%%%%%%%%%
%%%%%%%%%%%%%%%%%%%%%%%%%%%%%%%%%%%%%%%%%%%%%%%%%%%%%%%%%%%%%%%%%%%

\subsection{The discrete field}

We first describe the quantum probability 
space with which we model the electromagnetic 
field in a discrete manner. Imagine a 
one-dimensional field traveling towards 
a photo detector. We divide the field 
into $N$ time slices  of length $\lambda^2$. 
The total measurement time is $T = N\lambda^2$. 
If $N$ is large enough, the photo 
detector detects either zero or 
one photon in each time interval. 
Therefore, if $N$ is large, each slice of 
the field can in good approximation be 
regarded as a two-level system 
$({\cal M},\phi)$, see Example \ref{Quantum two-level system}. 
The {\it vacuum state} $\phi$ on ${\cal M}$
is given by $\phi(X)=\mean{\Phi,X\Phi}$ where 
$\Phi=[0~1]^{{\mathsf T}}$ denotes the so-called 
{\it vacuum vector}. 
The field can then be constructed as the $N$-fold tensor 
product of two-level systems representing 
the different time slices, i.e., $({\cal W}_N, \phi^{\otimes N})
=({\cal M}^{\otimes N}, \phi^{\otimes N})$. 
In particular, we assume that the system that 
interacts with the field is a two-level atomic 
system $({\cal M},\psi)$, i.e., the total 
quantum probability space for system and field 
together is given by 
\begin{equation}
\label{whole-space}
    ({\cal M}\otimes{\cal W}_N, {\mathbb P})
    =({\cal M}\otimes{\cal M}^{\otimes N}, \psi\otimes\phi^{\otimes N}). 
\end{equation}
Let $\rho$ be the density matrix corresponding to $\psi$. 
Then, ${\mathbb P}(X)$ can be written as 
${\mathbb P}(X)=\Tr[X(\rho\otimes(\Phi\Phi^*)^{\otimes N})]$ 
for all $X\in{\cal M}\otimes{\cal W}_N$.

Next, we introduce discrete noises. 
To this end, we define  
\[
    X_l:=I^{\otimes(l-1)}\otimes X\otimes I^{\otimes(N-l)}\in{\cal W}_N,~~
    l=1,\ldots,N, 
\]
where $X$ is a $2\times 2$ matrix and $I$ is 
the $2\times 2$ identity matrix. Using the 
above notation, let us define the following 
{\it noise matrices}: 
\begin{equation}
\label{noise}
    \Delta A(l)=\lambda(\sigma_-)_l,~~
    \Delta A(l)^*=\lambda(\sigma_+)_l,~~
    \Delta \Lambda(l)=(\sigma_+\sigma_-)_l,~~
    \Delta t(l)=\lambda^2 I_l, 
\end{equation}
where 
\begin{equation}
\label{creation-annihilation}
    \sigma_-=\left[ \begin{array}{cc}
               0 & 0 \\
               1 & 0 \\
             \end{array} \right],~~
    \sigma_+=\sigma_-^*
            =\left[ \begin{array}{cc}
               0 & 1 \\
               0 & 0 \\
             \end{array} \right]. 
\end{equation}
Furthermore, we define the following 
so-called {\it fundamental noises} living in the 
first $l$ slices: 
\[
    A(l)=\sum_{i=1}^l\Delta A(i),~~
    A(l)^*=\sum_{i=1}^l\Delta A(i)^*,~~
    \Lambda(l)=\sum_{i=1}^l\Delta \Lambda(i),~~
    t(l)=\sum_{i=1}^l\Delta t(i), 
\]
with the convention $A(0)= A(0)^* = \Lambda(0) = t(0)=0$.
We now provide the following physical interpretation 
to the fundamental noises. 
First, $t_l:=\iota(t(l))$ always takes 
the value $l\lambda^2=(l/N)T$, and thus, we may 
regard $t(l)$ as the time. 
Second, since 
$\Delta\lambda_l:=\iota(\Delta\Lambda(l))$ 
takes either the value $0$ 
or the value $1$ at time $l$, 
it is reasonable to interpret 
$\Lambda(l)$ as the total 
number of photons counted by the 
photo detector up to time $l$. 
For the vacuum state, we have 
${\rm Prob}(\Delta\lambda_l=1)=\phi^{\otimes N}({\rm diag}\{1,0\}_l)=0$, 
which implies that the photo detector detects no photons. 
Finally, with regard to $A(l)$ and $A(l)^*$, we 
introduce an observable 
$\Delta W(l):=\Delta A(l)+\Delta A(l)^*$ 
and a commutative $*$-algebra generated from 
$\Delta W(i),~(0\leq i\leq l)$: 
\begin{equation}
\label{c-algebra}
    {\cal C}_l
      :={\rm alg}\{\Delta W(i)=\Delta A(i)+\Delta A(i)^*~|~0\leq i\leq l\}. 
\end{equation}
$\Delta W(l)$ has the following spectral decomposition: 
\[
    \Delta W(l)=\Delta A(l)+\Delta A(l)^*
     =\left[ \begin{array}{cc}
           0 & \lambda \\
           \lambda & 0 \\
      \end{array} \right]_l
     =\lambda P_l^+ +(-\lambda) P_l^-, 
\]
with the projection matrices 
\begin{equation}
\label{random-walk-projector}
    P_l^+:=\frac{1}{2}
        \left[ \begin{array}{cc}
          1 & 1 \\
          1 & 1 \\
        \end{array} \right]_l,~~
    P_l^-:=\frac{1}{2}
        \left[ \begin{array}{cc}
          1 & -1 \\
          -1 & 1 \\
        \end{array} \right]_l. 
\end{equation}
Thus, for the vacuum state, the classical random variable 
$\Delta w_l:=\iota(\Delta W(l))$ takes $+\lambda$ with probability 
${\rm Prob}(+\lambda)=\phi^{\otimes N}(P_l^+)
=\mean{\Phi^{\otimes N},P_l^+\Phi^{\otimes N}}=1/2$ 
or $-\lambda$ with probability ${\rm Prob}(-\lambda)=1/2$ at each time. 
This implies that $\{w_l\}_{l=1,\ldots,N}$ is a 
symmetric random walk. If we let $N$ go to infinity and $\lambda$ 
to $0$, but keep the product $T=N\lambda^2$ constant, then 
it easily follows from Donsker's invariance principle 
(see e.g.\ \cite{kallenberg}) that $w_l$ converges weakly to a classical 
Brownian motion. Note 
that the relation $\Delta W(l)^2=\Delta t(l)$ 
becomes $dw_t^2=dt$ in the limit (see e.g.\ \cite{steele}). In physics the 
observable $A(l)+A(l)^*$ is known as a \emph{field 
quadrature}, see e.g.\ \cite{carmichael,gardiner}.

%%%%%%%%%%%%%%%%%%%%%%%%%%%%%%%%%%%%%%%%%%%%%%%%%%%%%%%%%%%%%%%%%%%
%%%%%%%%%%%%%%%%%%%%%%%%%%%%%%%%%%%%%%%%%%%%%%%%%%%%%%%%%%%%%%%%%%%

\subsection{System-field interaction}

Let ${\mathsf H}_1$ and ${\mathsf H}_2$ 
be Hilbert spaces with which we describe 
two separate quantum systems. The total interaction 
between these two systems over the first $l$ time 
units can be described by a unitary 
transformation $U(l)$ that acts on the composite space 
${\mathsf H}_1\otimes{\mathsf H}_2$. The time 
evolution of an observable $X$ of the composite 
system is given by the flow
  \begin{equation*}
  j_l(X) = U(l)^*XU(l).
  \end{equation*}
Suppose we start with an observable $X$ that 
acts non-trivially only on the first system. 
At time $l$ this observable is given 
by $j_l(X) = U(l)^*(X\otimes I)U(l)$ which 
in general will act non-trivially on both 
components in the tensor product 
${\mathsf H}_1\otimes {\mathsf H}_2$.  
This shows that information has been 
carried from the system that lives on ${\mathsf H}_1$ 
to the system on ${\mathsf H}_2$.  
Note that a unitary $U$ can always be represented 
as $U ={\rm e}^{-\im H}$ for 
some self-adjoint matrix $H$ 
called the {\it Hamiltonian}.

In our model, a two-level atomic system repeatedly 
interacts with the slices of the field. 
Let $H^{{\rm int}}(l)\in{\cal M}\otimes{\cal W}_l$
be the self-adjoint operator given by 
  \begin{equation}\label{hamiltonian}
    H^{{\rm int}}(l)
       =j_{l-1}(L_1)\otimes\Delta\Lambda(l)
                     +j_{l-1}(L_2)\otimes\Delta A(l)^*
                     +j_{l-1}(L_2^*)\otimes\Delta A(l)
                     +j_{l-1}(L_3)\otimes\Delta t(l), 
  \end{equation}
where the $L_i$'s are elements in ${\cal M}$ (for 
$i=1,2,3$) such that $L_1$ and $L_3$ are self-adjoint.
These system operators determine which kind of 
interaction between the two-level system and 
the field we are considering, i.e.\ they determine 
the physics of our problem. See Section 
\ref{sec examples} for two examples: a 
dispersive interaction and spontaneous decay.
We let $H^{{\rm int}}(l)$ be the Hamiltonian for the 
interaction between the system and the $l$-th field 
slice, that is, 
\begin{equation}
\label{repeated}
    U(l)
    =\mathop{\prod^{\longrightarrow}}_{i=1}^l
            {\rm e}^{-\im H^{{\rm int}}(i)}
    ={\rm e}^{-\im H^{{\rm int}}(1)}
            \cdots
            {\rm e}^{-\im H^{{\rm int}}(l)}, \qquad U(0) = I. 
\end{equation}  
We define another unitary operator $M_l$ by 
\begin{equation}
\label{hamiltonian-unitary}
    M_l:=
      {\rm e}^{ -\im\{
            L_1\otimes\Delta\Lambda(l)
            +L_2\otimes\Delta A^*(l)
            +L_2^*\otimes\Delta A(l)
            +L_3\otimes\Delta t(l) \}}. 
\end{equation}
Since ${\rm e}^{-\im H^{{\rm int}}(l)}=U(l-1)^*M_l U(l-1)$, 
the unitary operator $U(l)$ satisfies
\begin{equation}
\label{simple-discreteQsde}
    U(l)=U(l-1){\rm e}^{-\im H^{{\rm int}}(l)}=M_l U(l-1). 
\end{equation}
The operator $M_l$ acts non-identically only 
on the system and the $l$-th slice of the field. 
Thus, $M_l$ can be expressed as 
\begin{equation}
\label{M-evolution}
    M_l:=I+M^{\pm}\otimes\Delta\Lambda(l) + M^{+}\otimes\Delta A(l)^*
         + M^{-}\otimes\Delta A(l) + M^{\circ}\otimes\Delta t(l), 
\end{equation}
for some system operators $M^i\in{\cal M}~(i=\pm,+,-,\circ)$, 
which are uniquely determined by $L_i~(i=1,2,3)$. 
Note that the unitarity of $M_l$ implies certain 
relations between the operators $M^i$, e.g., 
$M^\circ + M^{\circ *} + M^{+*}M^+ + \lambda^2 M^{\circ *}M^\circ=0$.
From now on, we will use $M_l$ and 
$M^i$ instead of $H^{{\rm int}}(l)$ and 
$L_i$ to describe the interaction. 
We can write the following 
difference equation for the 
unitary $U(l)$  
\begin{equation}
\label{discreteQsde}
    \Delta U(l)=U(l)-U(l-1)
      =\big[
       M^{\pm}\Delta\Lambda(l) + M^{+}\Delta A(l)^*
       + M^{-}\Delta A(l) + M^{\circ}\Delta t(l) \big] U(l-1). 
\end{equation}
For simplicity we have omitted the tensor product 
$\otimes$ between $M^i$ and the noise operators. 
This rule will be applied throughout this paper. 
The dynamics (\ref{discreteQsde}) is called the 
{\it quantum stochastic difference equation}. It 
is a discrete version of the Hudson-Parthasarathy 
equation \cite{hudson}.

Next we describe a measurement performed 
on the field. Let us again consider the 
field observables $W(l)=A(l)+A(l)^*,~(l=0,\ldots,N)$. 
After the interaction, these observables 
are given by
\begin{equation}
\label{output-def}
    Y(l):=j_l(W(l))=U(l)^*\big[A(l)+A(l)^*\big]U(l),\qquad 0\le l\le N. 
\end{equation}
The observation process $Y_l,~(l=0,\ldots,N)$ 
satisfies the following difference equation
\begin{eqnarray}
\label{output-difference}
& & \hspace*{-1em}
    \Delta Y(l)
    =U(l)^*\big[\Delta A(l)+\Delta A(l)^*\big]U(l)
    =j_l(\Delta W(l)). 
\end{eqnarray}
Here we have used Eq.\ \eqref{simple-discreteQsde} and $[M_l, A(l-1)]=0$. 
Moreover, using $[M_k, \Delta W(l)]=0~(k\geq l+1)$ 
we find that 
\begin{eqnarray}
\label{yi-yj}
& & \hspace*{-1em}
    \Delta Y(l)
    =U(k)^*\Delta W(l)U(k)
    =j_k(\Delta W(l)), 
\end{eqnarray}
for all $k\geq l$. Therefore we find  
\begin{equation}
\label{self-QND}
    [\Delta Y(i),~\Delta Y(j)]=0,~~\forall i,j. 
\end{equation}
This means that the algebra generated by the observations
  \begin{equation}\label{eq observationalgebra}
 {\cal Y}_l={\rm alg}\{\Delta Y(i)~|~0\leq i\leq l\},
 \end{equation} 
is a commutative $*$-algebra for all $0\le l\le N$.
This is called the \emph{self-nondemolition} property
of the observations $Y_l$.  
Due to this property, we can define the classical process 
$\Delta y_l=\iota(\Delta Y(l)),~(l=0,\ldots,N)$. 
This classical process represents the data that 
we obtain while running the measurement. 
Note that $\Delta Y(l)$ has the following 
spectral decomposition 
\begin{equation}
\label{output-spectral}
    \Delta Y(l)
    =U(l)^*W(l)U(l)
    =\lambda U(l)^*P_l^+ U(l)+(-\lambda) U(l)^*P_l^- U(l), 
\end{equation}
where the projection matrices $P_l^+$ and $P_l^-$ are given by Eq.\ 
\eqref{random-walk-projector}. 
Hence, from Theorem~\ref{Spectral theorem}, the classical random variable 
$\Delta y_l=\iota(\Delta Y(l))$ takes $+\lambda$ with probability 
${\rm Prob}(+\lambda)=\psi\otimes\phi^{\otimes N}(U(l)^*P_l^+ U(l))$, 
which now depends on the interaction, or $-\lambda$ with probability 
${\rm Prob}(-\lambda)=1-{\rm Prob}(+\lambda)$.

%%%%%%%%%%%%%%%%%%%%%%%%%%%%%%%%%%%%%%%%%%%%%%%%%%%%%%%%%%%%%%%%%%%
%%%%%%%%%%%%%%%%%%%%%%%%%%%%%%%%%%%%%%%%%%%%%%%%%%%%%%%%%%%%%%%%%%%

\subsection{Quantum filtering}

The purpose of quantum filtering is to 
calculate the least mean square estimate 
of the observable $j_l(X)=U(l)^*XU(l)$ for 
a given system observable $X\in{\cal M}$. 
More specifically, we aim to find an element 
in the commutative $*$-algebra ${\cal Y}_l$ 
that minimizes the mean square error, i.e., 
$Z_l^{{\rm opt}}
=\mathop{\mathrm{argmin}}_{Z_l\in{\cal Y}_l}{\mathbb P}(|j_l(X)-Z_l|^2)$, 
where $|A|^2:=A^*A$ for an operator $A$ on a Hilbert space. 
Note that Eq.\ \eqref{yi-yj} leads to  
the following \emph{nondemolition} property 
\begin{equation}
\label{QND}
    [j_l(X),~\Delta Y(k)]=0,~~\forall N \ge l\geq k \ge 0, 
\end{equation}
which implies that $j_l(X)\in{\cal Y}_l',~\forall l$.
Due to the self-nondemolition and nondemolition properties 
the quantum conditional expectation 
${\mathbb{P}}(j_l(X)\condbar\mathcal{Y}_l)$ exists. 
Moreover, in Subsection \ref{sec condexp} we have seen 
that the quantum conditional expectation 
is an optimal estimator. Therefore the optimal 
estimator $Z_l^{{\rm opt}}$ is given by 
$Z_l^{{\rm opt}} = {\mathbb{P}}(j_l(X)\condbar\mathcal{Y}_l)$.

Since ${\mathbb{P}}(j_l(X)\condbar\mathcal{Y}_l)$ is linear, 
positive with respect to $X$, and normalized, i.e.\
${\mathbb{P}}(j_l(I)\condbar\mathcal{Y}_l) = 1$, we can define 
an \emph{information state} on the two-level atomic 
system by 
\begin{equation}
\label{def-of-pi}
  \pi_l(X) = {\mathbb{P}}(j_l(X)\condbar\mathcal{Y}_l),
  \qquad X \in \mathcal{M}.
\end{equation}
Note that the state $\pi_l$ on $\mathcal{M}$ is stochastic, 
it depends on the observations up to time $l$. 
We are now going to derive a difference equation 
for $\pi_l(X)$, i.e., the quantum filter. 
The following {\it noncommutative Bayes formula} 
\cite{luc3} is useful to derive the filter 
\begin{equation}\label{Q-kalli}
     \pi_l(X)
      ={\mathbb P}(j_l(X)\condbar{\cal Y}_l)
      =\frac
        {U(l)^*{\mathbb P}(V(l)^*XV(l)\condbar{\cal C}_l)U(l)}
        {U(l)^*{\mathbb P}(V(l)^*V(l)\condbar{\cal C}_l)U(l)}. 
\end{equation}
Here, ${\cal C}_l$ is the commutative $*$-algebra 
defined in Eq.\ \eqref{c-algebra} and $V(l)$ is the 
unique solution to the following difference equation: 
\begin{equation}
\label{standard-Veq}
     \Delta V(l)=\big[
       M^{+}\Delta W(l) + M^\circ\Delta t(l) \big] V(l-1),\qquad V(0)=I. 
\end{equation}
We note that the conditional expectation in Eq.\ \eqref{Q-kalli} is 
well defined, because $V(l)$ is driven by $\Delta t(l)$ and $\Delta W(l)$ and 
thus commutes with any element of ${\cal C}_l$. This means that 
$V(l)^*XV(l)$ is an element of ${\cal C}_l'$ for all 
system observables $X\in{\cal M}$. We now introduce an 
\emph{unnormalized information state} $\sigma_l$ by 
$\sigma_l(X) := U(l)^*{\mathbb P}(V(l)^*XV(l)\condbar{\cal C}_l)U(l)$ 
for all $X\in\mathcal{M}$. Eq.\ \eqref{Q-kalli} now reads 
$\pi_l(X) = \sigma_l(X)/\sigma_l(I)$, which is a noncommutative 
analogue of the classical \emph{Kallianpur-Striebel formula} \cite{kallianpurstrebel}. 
It easily follows from Eqs.\ (\ref{discreteQsde}) and 
(\ref{standard-Veq}) that $\sigma_l(X)$ satisfies the 
following difference equation  
\begin{equation}
\label{Q-zakai}
    \Delta \sigma_l(X)
       =\sigma_{l-1}({\cal L}(X))\Delta t(l)
       +\sigma_{l-1}({\cal J}(X))\Delta Y(l),~~~
    \sigma_0=\psi, 
\end{equation}
where the operators ${\cal L}$ and ${\cal J}$ are given by 
\begin{eqnarray}
\label{lindbladian}
& & \hspace*{0em}
    {\cal L}(X)
        :=M^{+*}XM^+ + \lambda^2 M^{\circ *}XM^\circ 
           + XM^\circ + M^{\circ *}X, 
\nonumber \\ & & \hspace*{0em}
    {\cal J}(X)
        :=\lambda^2 M^{+*}XM^\circ + \lambda^2 M^{\circ *}XM^+ 
          + XM^+ + M^{+*}X. 
\end{eqnarray}
The filter can now be obtained immediately 
from $\pi_l(X) = \sigma_l(X)/\sigma_l(I)$.
We, however, will always use the unnormalized
version of the filter given in Eq.\ \eqref{Q-zakai}.

Note that $\sigma_l(X)$, $\Delta t(l)$, 
and $\Delta Y(l)$ are all elements in 
the commutative $*$-algebra ${\cal Y}_l$. 
Due to Theorem \ref{Spectral theorem} they can be diagonalized 
simultaneously, which yields the following 
classical random variables 
$\iota(\sigma_l(X))$, $\Delta t_l=\iota(\Delta t(l))=\lambda^2$, and 
$\Delta y_l=\iota(\Delta Y(l))$. 
Moreover, since $\iota(\sigma_l(X))$ is a 
linear and positive functional of $X$, 
we can define a $2\times 2$ positive 
semidefinite matrix $\varrho_l$ that 
satisfies $\iota(\sigma_l(X))=\Tr(\varrho_l X)$. 
The unnormalized density matrix $\varrho_l$ is  
called the {\it unnormalized information density matrix}. 
It is easy to derive a difference 
equation for $\varrho_l$: 
\begin{equation}
\label{Q-zakai-infor-state}
    \Delta \varrho_l
       =\bar{{\cal L}}(\varrho_{l-1})\Delta t_l
       +\bar{{\cal J}}(\varrho_{l-1})\Delta y_l, ~~~
    \varrho_0=\rho, 
\end{equation}
where the operators $\bar{{\cal L}}$ and $\bar{{\cal J}}$ are 
given by 

\begin{eqnarray}
& & \hspace*{0em}
\label{def-of-Lbar}
    \bar{{\cal L}}(\varrho)
        :=M^+ \varrho M^{+*} + \lambda^2 M^\circ\varrho M^{\circ *}
        + M^\circ \varrho + \varrho M^{\circ *}, 
\\ & & \hspace*{0em}
\label{def-of-Jbar}
    \bar{{\cal J}}(\varrho)
        :=\lambda^2 M^\circ \varrho M^{+*} + \lambda^2 M^+ \varrho M^{\circ *} 
         + M^+\varrho + \varrho M^{+*}. 
\end{eqnarray}
%

%%%%%%%%%%%%%%%%%%%%%%%%%%%%%%%%%%%%%%%%%%%%%%%%%%%%%%%%%%%%%%%%%%%
%%%%%%%%%%%%%%%%%%%%%%%%%% The composition %%%%%%%%%%%%%%%%%%%%%%%%
%%%%%%%%%%%%%%%%%%%%%%%%%%%%%%%%%%%%%%%%%%%%%%%%%%%%%%%%%%%%%%%%%%%

\section{Composition of an operator-valued function and an observable}

In the following section we will formulate 
risk sensitive estimation as an optimal control 
problem for a given cost function, see Eqs.\ 
\eqref{def-of-rs-estimator} and \eqref{costfunction} below. The idea of 
risk-sensitive control is to absorb the running 
cost of the cost function into the dynamics, 
see Eq.\ \eqref{def-of-Umu} below. This means 
that the new dynamics depends on past 
estimates (the controls in the optimal control 
formulation of risk-sensitive estimation) which 
are a function of the observations thus far.
Therefore we need to make precise mathematically what 
we mean by operator coefficients (see for example the 
coefficients in Eq.\ \eqref{dynamics-risksensitive} 
below) that depend on a function of the 
observations thus far. We address this topic 
in this section.

Let ${\cal A}_1$  be a finite dimensional  
$*$-algebra and let ${\cal A}_2$ 
be a commutative finite dimensional $*$-algebra.
Let $K$ be an ${\cal A}_1$-valued function on 
$\mathbb{C}$, i.e., 
$K:{\mathbb C}\ni u\rightarrow K(u)\in{\cal A}_1$. 
Let ${\mathfrak u}$ be an
element in ${\cal A}_2$.  
Note that since $\mathcal{A}_2$ is commutative, we have
${\mathfrak u}^*{\mathfrak u}={\mathfrak u}{\mathfrak u}^*$, 
i.e., ${\mathfrak u}$ is \emph{normal}. 
The spectral decomposition of $\mathfrak{u}$  
can be written as
${\mathfrak u}=\sum_{x\in{\rm sp}({\mathfrak u})}x P_{{\mathfrak u}}(x)$,
where ${\rm sp}({\mathfrak u})$ denotes 
the spectrum of ${\mathfrak u}$, i.e., 
the set of eigenvalues of $\mathfrak{u}$. 
The composition $K({\mathfrak u})\in{\cal A}_1\otimes{\cal A}_2$ 
of $K$ with $\mathfrak{u}$ is 
defined as \cite{luc2}
\begin{equation}
\label{composition}
    K({\mathfrak u})
      :=\sum_{x\in{\rm sp}({\mathfrak u})}
         K(x) P_{{\mathfrak u}}(x)
            \in{\cal A}_1\otimes{\cal A}_2. 
\end{equation}
This is a natural generalization of the 
composition of $K$ with a classical 
random variable $\alpha:\ 
(\Omega,{\mathcal F},{\bf P}) \to \mathbb{C}$,  
given by 
\begin{equation}
\label{classical-composition}
    K(\alpha)(\omega)
      :=\sum_{x\in{\rm ran}(\alpha)}
         K(x)\chi_{\{\alpha=x\}}(\omega)
    \in{\mathcal A}_1. 
\end{equation}
Here ${\rm ran}(\alpha)$ denotes the range 
of $\alpha$ and $\chi_{\{\alpha=x\}}$ 
is the indicator function of the set 
$\{\omega\in\Omega\condbar \alpha(\omega)=x\}$.

Let ${\mathfrak u}_l$ be an element of the 
observation algebra
${\mathcal Y}_l$, defined in 
Eq.\ \eqref{eq observationalgebra}. 
This means that we can 
write ${\mathfrak u}_l$ as a function of 
$\Delta Y(i),~1\le i\le l$: 
\[
    {\mathfrak u}_l
      =f_l(\Delta Y(1),\ldots,\Delta Y(l))\in{\mathcal Y}_{l}, 
\]
for some function 
$f_l:{\mathbb R}^{l}\rightarrow{\mathbb C}$.
Moreover, we can also write ${\mathfrak u}_l$ 
in terms of the observables 
$\Delta W(i)=\Delta A(i)+\Delta A(i)^*,~1\le i\le l$ 
as
\[
    {\mathfrak u}_l
    =f_l( j_{l}(\Delta W(1)), \ldots, j_{l}(\Delta W(l)) )
    =j_{l}(f_l(\Delta W(1),\ldots,\Delta W(l))), 
\]
where we have used Eq.\ \eqref{yi-yj}. 
Therefore, if we define an element 
${\check {\mathfrak u}}_l$ in ${\cal C}_l$ by 
\[
    {\check {\mathfrak u}}_l
     :=f_l(\Delta W(1),\ldots,\Delta W(l))\in{\mathcal C}_{l},
\]
then ${\mathfrak u}_l$ can be written as 
${\mathfrak u}_l=j_{l}({\check {\mathfrak u}}_l)$. 
An ${\cal M}$-valued function 
$K: {\mathbb C}\ni u\rightarrow K(u)\in{\cal M}$ and an element 
${\check {\mathfrak u}}_l$ in ${\cal C}_l$ 
give rise to the composition 
$K({\check {\mathfrak u}}_l)$, which is an 
element in ${\cal M}\otimes{\cal C}_l$. 
Denoting the spectral decomposition of ${\check {\mathfrak u}}_l$ as 
${\check {\mathfrak u}}_l=\sum_{x\in {\rm sp}(\check{\mathfrak{u}}_l)} 
x P_{{\check {\mathfrak u}}_l}(x)$, 
we obtain 
\begin{eqnarray}
\label{Kl-K(l)}
& & \hspace*{0em}
    j_{l}(K({\check {\mathfrak u}}_l))
    =\sum_{x\in {\rm sp}(\check{\mathfrak{u}}_l)} 
    U(l)^* K(x) U(l)U(l)^*P_{{\check {\mathfrak u}}_l}(x)U(l)
    =\sum_{x\in {\rm sp}(\check{\mathfrak{u}}_l)} 
    j_l(K(x)) P_{U(l)^*{\check {\mathfrak u}}_lU(l)}(x)
\nonumber \\ & & \hspace*{4.3em}
    =\sum_{x\in {\rm sp}(\mathfrak{u}_l)} 
    j_l(K(x))P_{{\mathfrak u}_l}(x)=: j_l(K)({\mathfrak u}_l), 
\end{eqnarray}
where we have introduced the notation $j_l(K)({\mathfrak u}_l)$
in the last step. Note that $j_l(K)({\mathfrak u}_l)$ is 
an element in $U(l)^*({\cal M}\otimes{\cal C}_l)U(l)$.

%%%%%%%%%%%%%%%%%%%%%%%%%%%%%%%%%%%%%%%%%%%%%%%%%%%%%%%%%%%%%%%%%%%
%%%%%%%%%%%%%%%%%%% Q risk-sensitive filtering %%%%%%%%%%%%%%%%%%%%
%%%%%%%%%%%%%%%%%%%%%%%%%%%%%%%%%%%%%%%%%%%%%%%%%%%%%%%%%%%%%%%%%%%

\section{Quantum risk-sensitive filtering}

In this section we study a quantum risk-sensitive 
estimation problem. Let $X_{{\rm e}}$ be a fixed element of 
the two-level atomic system $\mathcal{M}$. 
Then, the risk-sensitive estimator of $j_l(X_{{\rm e}})$ is defined as 
follows: 
\begin{equation}
\label{def-of-rs-estimator}
    \left(\widehat{X_{{\rm e}}}^\mu(1),\ldots,
                         \widehat{X_{{\rm e}}}^\mu(N)\right)
        :=\mathop{\mathrm{argmin}}_{
           {\mathfrak u}_1\in{\cal Y}_1,\ldots,{\mathfrak u}_N\in{\cal Y}_N}
             F({\mathfrak u}_1,\ldots,{\mathfrak u}_N),
\end{equation}	     
where the {\it cost function} $F$ is given by	     
\begin{equation}\label{costfunction}
    F({\mathfrak u}_1,\ldots,{\mathfrak u}_N):=
              {\mathbb P}\Big[
      R(N)^*{\rm exp}\Big(\mu_2|j_N(X_{{\rm e}})
                            -{\mathfrak u}_N|^2\Big) R(N) \Big], 
\end{equation}
and the matrix 
$R(l)\in{\cal M}\otimes{\cal W}_{l-1}$ 
is given by 
\begin{equation}
\label{def-of-R}
    R(l)
    =\mathop{\prod^{\longleftarrow}}_{i=1}^{l-1}
       {\rm exp}\Big[ \frac{\mu_1}{2}\lambda^2 
          |j_i(X_{{\rm e}})-{\mathfrak u}_i|^2 \Big],~~
    R(1)=R(0)=I. 
\end{equation}
Note again that $|A|^2:=A^*A$. 
Here, $\mu = (\mu_1,\mu_2)$ are weighting 
parameters that represent risk-sensitivity. 
Using the ${\cal M}$-valued function 
\[
    K:\ {\mathbb C} \to {\cal M}:\ u\rightarrow K(u)=|X_{{\rm e}}-u|^2, 
\]
we can write 
$K({\check {\mathfrak u}}_l)=|X_{{\rm e}}-{\check {\mathfrak u}}_l|^2$ 
and $j_l(K)({\mathfrak u}_l)=|j_l(X_{{\rm e}})-{\mathfrak u}_l|^2$.  
Using these compositions, we can obtain a recursive form of $R(l)$: 
\begin{eqnarray}
\label{simple-rep-R}
& & \hspace*{0em}
    R(l)
    =\mathop{\prod^{\longleftarrow}}_{i=1}^{l-1}
       {\rm exp}\Big[ \frac{\mu_1}{2}\lambda^2 j_i(K)({\mathfrak u}_i) \Big]
    =\mathop{\prod^{\longleftarrow}}_{i=1}^{l-1}
       {\rm exp}\Big[ \frac{\mu_1}{2}\lambda^2 
          j_i(K({\check {\mathfrak u}}_i)) \Big]
\nonumber \\ & & \hspace*{2.12em}
    ={\rm exp}\Big[ \frac{\mu_1}{2}\lambda^2 
          j_{l-1}(K({\check {\mathfrak u}}_{l-1})) \Big]R(l-1)
    =j_{l-1}\Big(
         {\rm e}^{\mu_1\lambda^2 K({\check {\mathfrak u}}_{l-1})/2}\Big)
             R(l-1). 
\end{eqnarray}

\begin{remark}
\label{RS-cost in commutative case}
If all matrices in Eqs.\ \eqref{costfunction} and \eqref{def-of-R} 
commute with each other, the quantum risk-sensitive estimator reduces to 
\[
    \left(\widehat{X_{{\rm e}}}^\mu(1),\ldots,
           \widehat{X_{{\rm e}}}^\mu(N)\right)=
      \mathop{\mathrm{argmin}}_{
             {\mathfrak u}_1\in{\cal Y}_1,\ldots,{\mathfrak u}_N\in{\cal Y}_N}
          {\mathbb P}\Big[
             {\rm exp}\Big(
                \mu_1\lambda^2\sum_{i=1}^{N-1}|j_i(X_{{\rm e}})
                                                 -{\mathfrak u}_i|^2
                +\mu_2|j_N(X_{{\rm e}})-{\mathfrak u}_N|^2 \Big)\Big], 
\]
which is identical to the definition of 
the (generalized) classical risk-sensitive 
estimator in Eq.\ \eqref{classical-rs-estimator}. 
Hence, Eq.\ \eqref{def-of-rs-estimator} is a natural 
noncommutative extension of the classical 
risk-sensitive estimator to the quantum case. 
\end{remark}

\begin{remark}
\label{Risk-netral}
In the limit of $\mu_1,\mu_2\rightarrow 0$, $\widehat{X_{{\rm e}}}^\mu(l)$ 
coincides with the standard quantum optimal estimator $\pi_l(X)$ in 
Eq.\ \eqref{def-of-pi}. 
This is easily seen as follows. 
The estimation error cost function 
in Eq.\ \eqref{costfunction} is 
expanded to first order in $\mu_1$ and $\mu_2$ as 
\[
     F({\mathfrak u}_1,\ldots,{\mathfrak u}_N)=
       1+\mu_1\lambda^2\sum_{i=1}^{N-1}
           {\mathbb P}\big(|j_i(X_{{\rm e}})-{\mathfrak u}_i|^2\big)
        +\mu_2{\mathbb P}\big(|j_N(X_{{\rm e}})-{\mathfrak u}_N|^2\big)
        +o(\mu_1,\mu_2). 
\]
Thus, in the limit $\mu_1,\mu_2\rightarrow 0$, the minimizers of this 
function are given by ${\mathfrak u}^{{\rm opt}}_l=\pi_l(X_{{\rm e}})$, 
$(l=1,\ldots,N)$, i.e., we have 
\begin{equation}
\label{small-risk-filtering}
     \lim_{\mu_1,\mu_2\rightarrow 0}\widehat{X_{{\rm e}}}^\mu(l)
              =\pi_l(X_{{\rm e}}). 
\end{equation}
For this reason, $\pi_l(X_{{\rm e}})$ is called 
the risk-neutral estimator. 
\end{remark}

The remainder of this section is organized as follows. 
First, we introduce a risk-sensitive information 
density matrix $\varrho^\mu_l$, which is the 
quantum analogue to the classical information state 
$\alpha_l^\mu(x)$ discussed in Section I-A. 
Second, we derive a recursive equation for $\varrho^\mu_l$. 
As in the standard quantum filtering case, $\varrho^\mu_l$ contains all 
information needed to calculate the estimator \eqref{def-of-rs-estimator}. 
More specifically, Eq.\ \eqref{costfunction} 
can be represented only in terms of $\varrho^\mu_l$, 
see Section V-C.

%%%%%%%%%%%%%%%%%%%%%%%%%%%%%%%%%%%%%%%%%%%%%%%%%%%%%%%%%%%%%%%%%%%
%%%%%%%%%%%%%%%%%%%%%%%%%%%%%%%%%%%%%%%%%%%%%%%%%%%%%%%%%%%%%%%%%%%

\subsection{Quantum risk-sensitive information state}

We start by defining the following modification 
of the unitaries given by the difference 
equation~\eqref{discreteQsde}: 
\begin{equation}
\label{def-of-Umu}
    U^\mu(l):=U(l) R(l). 
\end{equation}
Here $R(l)$ is given by Eq.\ \eqref{def-of-R}.
Note that $R(l)$ depends on $\mu = (\mu_1,\mu_2)$, but 
only through $\mu_1$. 
Using Eqs. \eqref{simple-discreteQsde} 
and \eqref{simple-rep-R}, we 
find the following difference equation 
for $U^\mu(l)$
\begin{eqnarray}\label{dynamics-risksensitive}
& & \hspace*{-1em}
    \Delta U^\mu(l)= U(l) R(l)-U^\mu(l-1)
\nonumber \\ & & \hspace*{2.52em}
    =M_l U(l-1)
     U(l-1)^*{\rm e}^{\mu_1\lambda^2 K_{l-1}/2}U(l-1)R(l-1)
     -U^\mu(l-1)
\nonumber \\ & & \hspace*{2.52em}
    =\Big[ M_l {\rm e}^{\mu_1\lambda^2 K_{l-1}/2}-I\Big]U^\mu(l-1)
    ,\qquad U^\mu(0) = I. 
\nonumber
\end{eqnarray}
Here, we have used $K_{l-1}$ as a short hand for 
$K({\check {\mathfrak u}}_{l-1})$. 
Using Eq.\ \eqref{M-evolution}, this can be rewritten as
\begin{eqnarray}
& & \hspace*{-1em}
    \Delta U^\mu(l)=\Big[
      \Big\{ M^\circ{\rm e}^{\mu_1\lambda^2 K_{l-1}/2} 
           + \frac{1}{\lambda^2}({\rm e}^{\mu_1\lambda^2 K_{l-1}/2}-1) \Big\}
             \Delta t(l)
\nonumber \\ & & \hspace*{-1em}
    \mbox{}
    +M^{\pm}{\rm e}^{\mu_1\lambda^2 K_{l-1}/2}\Delta\Lambda(l)
    +M^{+}{\rm e}^{\mu_1\lambda^2 K_{l-1}/2}\Delta A(l)^*
    +M^{-}{\rm e}^{\mu_1\lambda^2 K_{l-1}/2}\Delta A(l) \Big]U^\mu(l-1). 
\nonumber
\end{eqnarray}
Now, let us define $V^\mu(l)$ as 
the solution to the following difference 
equation 
\begin{equation}
\label{Vtilde-eq1}
    \Delta V^\mu(l)=\Big[
      \Big\{ M^\circ{\rm e}^{\mu_1\lambda^2 K_{l-1}/2} 
           + \frac{1}{\lambda^2}({\rm e}^{\mu_1\lambda^2 K_{l-1}/2}-1) \Big\}
             \Delta t(l)
    +M^{+}{\rm e}^{\mu_1\lambda^2 K_{l-1}/2}\Delta W(l)
        \Big]V^\mu(l-1), 
\end{equation}
with $V^\mu(0)=I$. 
Note that this equation is identical to Eq.\ \eqref{standard-Veq} when 
$\mu_1=0$. 
Two crucial properties of $V^\mu(l)$ are given in the following lemma.

\begin{lemma}
\label{U-V connection}
For all $1\le l \le N$ the matrix $V^\mu(l)$ 
is an element of 
${\cal M}\otimes{\cal C}_l\subset{\cal C}_l'$. 
Moreover, we have
\begin{equation}
\label{V-tildeV}
    {\mathbb P}\Big[U^\mu(l)^*X U^\mu(l)\Big]
      ={\mathbb P}\Big[V^\mu(l)^*XV^\mu(l)\Big],
\end{equation}
for any $X$ in ${\cal M}\otimes{\cal W}_N$. 
\end{lemma}

\begin{proof}
To prove the first assertion, we assume that 
$V^\mu(l-1)\in{\cal M}\otimes{\cal C}_{l-1}$. 
Since $V^\mu(l)$ is calculated recursively, 
using $\Delta W(l)$, $\Delta t(l)$, 
and $V^\mu(l-1)$, all of which are included in 
${\cal M}\otimes{\cal C}_l$, we obtain 
$V^\mu(l)\in{\cal M}\otimes{\cal C}_l$. 
The assertion follows by induction.

For the second claim, we note that 
$U^\mu(l)v\otimes\Phi^{\otimes N}=V^\mu(l)v\otimes\Phi^{\otimes N}$ 
holds for all vectors $v\in{\mathbb C}^2$ due to the relations 
$\Delta A(l)\Phi^{\otimes N}=\Delta \Lambda(l)\Phi^{\otimes N}=0$ 
and $U^\mu(0)=V^\mu(0)=I$. 
Therefore, when the system density matrix is of the form $\rho=vv^*$, 
any $X\in{\cal M}\otimes{\cal W}_N$ satisfies
\[
     \left\langle U^\mu(l)v\otimes\Phi^{\otimes N}, 
                     XU^\mu(l)v\otimes\Phi^{\otimes N} \right\rangle
     = \left\langle V^\mu(l)v\otimes\Phi^{\otimes N}, 
                   XV^\mu(l)v\otimes\Phi^{\otimes N} \right\rangle, 
\]
which directly implies Eq.\ \eqref{V-tildeV} 
due to Eq.\ \eqref{pure-state}. 
Since every density matrix $\rho$ is a convex 
combination of vector states, the lemma is proved. 
\end{proof}

\begin{definition}
Since by Lemma \ref{U-V connection} $V^\mu(l)$ is an 
element of the commutant of $\mathcal{C}_l$, 
we can define the following unnormalized risk-sensitive 
information state \cite{dhelon}: 
  \begin{equation}
  \label{def-of-sigma}
     \sigma_l^\mu(X)
      :=U(l)^*{\mathbb P}(V^\mu(l)^*XV^\mu(l)\condbar{\cal C}_l)U(l)
         \in{\cal Y}_l. 
  \end{equation}
Moreover, we define $\varrho_l^\mu$ as the 
unnormalized risk-sensitive information density 
matrix corresponding to $\sigma^\mu_l$ by 
  \begin{equation}
  \label{def-rs-info-state}
       \iota(\sigma_l^\mu(X))=\Tr(\varrho_l^\mu X),~~\forall X\in{\cal M}.
  \end{equation}
\end{definition}
\mbox{}

\begin{lemma}
\label{info-composition}
Let ${\check {\mathfrak u}}_l$ be an element 
in ${\cal C}_l$. Let $Z: \mathbb{C} \to 
{\cal M}$ be an ${\cal M}$-valued 
function. Then we have  
\begin{equation*}
\label{def-rs-info-state-final}
      \iota\Big(\sigma^\mu_l(Z({\check {\mathfrak u}}_l))\Big)
        =\Tr\big[\varrho^\mu_l Z(u_l) \big],
\end{equation*}
where 
$u_l=\iota(j_l({\check {\mathfrak u}}_l))=\iota({\mathfrak u}_l)$ 
is a function of $\Delta y_i=\iota(\Delta Y(i)),~1\leq i\leq l$. 
\end{lemma}

\begin{proof}
Denote the spectral decomposition of ${\check {\mathfrak u}}_l$ by 
${\check {\mathfrak u}}_l
=\sum_{x \in {\rm sp}({\check {\mathfrak u}}_l)} 
xP_{{\check {\mathfrak u}}_l}(x)\in{\cal C}_l$. 
Then, it follows from the definitions (\ref{composition}) and 
(\ref{def-of-sigma}) that we have 
\begin{eqnarray}
& & \hspace*{-1em}
    \sigma_l^\mu\big(Z({\check {\mathfrak u}}_l)\big)
     =U(l)^*{\mathbb P}\Big[
        V^\mu(l)^*\Big( \sum_{x \in {\rm sp}({\check {\mathfrak u}}_l)}
            Z(x)P_{{\check {\mathfrak u}}_l}(x)\Big)V^\mu(l)
           \condbigbar{\cal C}_l\Big]U(l)
\nonumber \\ & & \hspace*{3.7em}
    =\sum_{x \in {\rm sp}({\check {\mathfrak u}}_l)}
       U(l)^*{\mathbb P}\Big(
               V^\mu(l)^* Z(x) V^\mu(l)\condbigbar{\cal C}_l\Big)U(l)
                  U(l)^*P_{{\check {\mathfrak u}}_l}(x)U(l)
\nonumber \\ & & \hspace*{3.7em}
    =\sum_{x \in {\rm sp}({\check {\mathfrak u}}_l)}
       \sigma_l^\mu(Z(x)) P_{U(l)^*{\check {\mathfrak u}}_lU(l)}(x)
    =\sum_{x \in {\rm sp}({\mathfrak u}_l)}
       \sigma_l^\mu(Z(x)) P_{{\mathfrak u}_l}(x). 
\nonumber
\end{eqnarray}
In the first step we used 
$P_{{\check {\mathfrak u}}_l}(x)\in{\cal C}_l$ 
and $[P_{{\check {\mathfrak u}}_l}(x), V^\mu(l)]=0$. 
Note that $\sigma_l^\mu(Z(x))\in{\cal Y}_l$ and 
$P_{{\mathfrak u}_l}(x)\in{\cal Y}_l$ can be diagonalized 
simultaneously by a $*$-isomorphism $\iota$. 
Using $\iota(P_{{\mathfrak u}_l}(x))
=\chi_{\{\iota({\mathfrak u}_l)=x\}}$ 
(see Theorem \ref{Spectral theorem}), we get
\begin{eqnarray}
& & \hspace*{-1em}
    \iota\Big(\sigma_l^\mu\big(Z({\check {\mathfrak u}}_l)\big)\Big)
     =\sum_{x \in {\rm sp}({\mathfrak u}_l)}
        \iota\Big(\sigma_l^\mu(Z(x))\Big) 
             \iota\Big(P_{{\mathfrak u}_l}(x)\Big)
    =\sum_{x \in {\rm sp}({\mathfrak u}_l)}
        \Tr\Big(\varrho_l^\mu Z(x)\Big)
            \chi_{\{\iota({\mathfrak u}_l)=x\}}
\nonumber \\ & & \hspace*{5.2em}
    =\Tr\Big[\varrho_l^\mu 
        \sum_{x \in {\rm ran}(u_l)}
           Z(x)\chi_{\{u_l=x\}}\Big]
    =\Tr\Big[\varrho_l^\mu Z(u_l)\Big], 
\nonumber
\end{eqnarray}
where we have used the definitions (\ref{classical-composition}) and 
(\ref{def-rs-info-state}). 
Since ${\mathfrak u}_l\in{\cal Y}_l$, $u_l=\iota({\mathfrak u}_l)$ is 
obviously a function of $\Delta y_1,\ldots,\Delta y_l$. 
This completes the proof. 
\end{proof}

%%%%%%%%%%%%%%%%%%%%%%%%%%%%%%%%%%%%%%%%%%%%%%%%%%%%%%%%%%%%%%%%%%%
%%%%%%%%%%%%%%%%%%%%%%%%%%%%%%%%%%%%%%%%%%%%%%%%%%%%%%%%%%%%%%%%%%%

\subsection{Dynamics of risk-sensitive information density matrix}

The objective here is to derive a 
recursive equation for $\varrho^\mu_l$. 
Let $X$ be an element of ${\cal M}$.
A similar calculation 
to Eq.\ \eqref{output-difference} yields  
the following difference equation for 
$\tilde{j}^\mu_l(X):=V^\mu(l)^*XV^\mu(l)$
\[
    \Delta \tilde{j}^\mu_l(X)
      =\tilde{j}^\mu_{l-1}({\cal L}^\mu
         (X,{\check {\mathfrak u}}_{l-1}))\Delta t(l)
      +\tilde{j}^\mu_{l-1}
         ({\cal J}^\mu(X,{\check {\mathfrak u}}_{l-1}))\Delta W(l), 
\]
where 
\begin{eqnarray}
& & \hspace*{-1em}
    {\cal L}^\mu(X,u)
        :={\rm e}^{\mu_1\lambda^2 K(u)/2}
            \Big[ M^{+*}XM^+ + \lambda^2 M^{o*}XM^o 
     + XM^o + M^{o*}X \Big]
                        {\rm e}^{\mu_1\lambda^2 K(u)/2}
\nonumber \\ & & \hspace*{7em}
     \mbox{}
     +\frac{1}{\lambda^2}\Big(
         {\rm e}^{\mu_1\lambda^2 K(u)/2}
              X{\rm e}^{\mu_1\lambda^2 K(u)/2}-X \Big)\in{\cal M}, 
\nonumber \\ & & \hspace*{-1em}
    {\cal J}^\mu(X,u)
        :={\rm e}^{\mu_1\lambda^2 K(u)/2}
            \Big[ \lambda^2 M^{+*}XM^o + \lambda^2 M^{o*}XM^+ 
\nonumber \\ & & \hspace*{7em}
    \mbox{}
       + XM^+ + M^{+*}X\Big]
            {\rm e}^{\mu_1\lambda^2 K(u)/2}\in{\cal M}. 
\nonumber
\end{eqnarray}
Note that $K(u)=|X_{{\rm e}}-u|^2$. 
Since $\tilde{j}^\mu_l(X)$ is an element 
of the commutant ${\cal C}'_l$, we can 
define the quantum conditional expectation 
$\tilde{\sigma}^\mu_l(X):={\mathbb P}(\tilde{j}^\mu_l(X)\condbar{\cal C}_l)$. 
This satisfies the following difference equation  
\[
    \Delta \tilde{\sigma}^\mu_l(X)
      =\tilde{\sigma}^\mu_{l-1}
        ({\cal L}^\mu(X,{\check {\mathfrak u}}_{l-1}))\Delta t(l)
      +\tilde{\sigma}^\mu_{l-1}
        ({\cal J}^\mu(X,{\check {\mathfrak u}}_{l-1}))\Delta W(l). 
\]
Eq.\ \eqref{def-of-sigma} can now be written as 
$\sigma^\mu_l(X)=U(l)^*\tilde{\sigma}^\mu_l(X)U(l)$. 
This means we find the following difference equation  
\begin{eqnarray}
\label{risk-zakai}
& & \hspace*{-1em}
    \Delta \sigma^\mu_l(X)
      =U(l-1)^* M_l^*\Delta\tilde{\sigma}^\mu_l(X)M_l U(l-1)
\nonumber \\ & & \hspace*{2.9em}
    =U(l-1)^*\tilde{\sigma}^\mu_{l-1}
            ({\cal L}^\mu(X,{\check {\mathfrak u}}_{l-1}))U(l-1)\Delta t(l)
\nonumber \\ & & \hspace*{5em}
    \mbox{}
    +U(l-1)^*\tilde{\sigma}^\mu_{l-1}
            ({\cal J}^\mu(X,{\check {\mathfrak u}}_{l-1}))U(l-1)
       U(l)^*\Delta W(l)U(l)
\nonumber \\ & & \hspace*{2.9em}
    =\sigma^\mu_{l-1}
        ({\cal L}^\mu(X,{\check {\mathfrak u}}_{l-1}))\Delta t(l)
    +\sigma^\mu_{l-1}
        ({\cal J}^\mu(X,{\check {\mathfrak u}}_{l-1}))\Delta Y(l), 
\end{eqnarray}
where in the last step Eq.\ \eqref{output-difference} 
was used.

We can now represent Eq.\ \eqref{risk-zakai} 
in terms of the unnormalized risk-sensitive 
information density matrix $\varrho_l^\mu$. 
Since $\sigma^\mu_l(X),\ \Delta t(l)$, and $\Delta Y(l)$ 
are elements in ${\cal Y}_l$, they can be simultaneously 
diagonalized by a $*$-isomorphism $\iota$, which leads to 
\[
    \Delta \iota(\sigma^\mu_l(X))
       =\iota(\sigma^\mu_{l-1}({\cal L}^\mu
               (X,{\check {\mathfrak u}}_{l-1})))\Delta t_l
       +\iota(\sigma^\mu_{l-1}
               ({\cal J}^\mu(X,{\check {\mathfrak u}}_{l-1})))\Delta y_l, 
	       \qquad \sigma_0^\mu = \psi. 
\]
where $\Delta t_l=\iota(\Delta t(l))$ and 
$\Delta y_l=\iota(\Delta Y(l))$. It then follows 
from Lemma \ref{info-composition} that the above 
equation leads to 
\begin{equation}
\label{zakai-SME}
    \Delta \varrho^\mu_l
       =\bar{{\cal L}}^\mu(\varrho^\mu_{l-1},u_{l-1})\Delta t_l
       +\bar{{\cal J}}^\mu(\varrho^\mu_{l-1},u_{l-1})\Delta y_l, 
       \qquad \varrho^\mu_0 = \rho.
\end{equation}
where $u_{l-1}=\iota(j_{l-1}({\check {\mathfrak u}}_{l-1}))
=\iota({\mathfrak u}_{l-1})$ 
is a function of $\Delta y_1,\ldots,\Delta y_{l-1}$. 
The operators $\bar{{\cal L}}^\mu$ and $\bar{{\cal J}}^\mu$ are defined 
as follows 
\begin{eqnarray}
& & \hspace*{-1em}
    \bar{{\cal L}}^\mu
      (\varrho,u)
        :=M^+{\cal H}(\varrho,u) M^{+*} 
        + \lambda^2 M^\circ{\cal H}(\varrho,u) M^{\circ*} 
\nonumber \\ & & \hspace*{5em}
     \mbox{}
        + M^\circ{\cal H}(\varrho,u) + {\cal H}(\varrho,u) M^{\circ*}
        +\frac{1}{\lambda^2}({\cal H}(\varrho,u)-\varrho),
\nonumber \\ & & \hspace*{-1em}
    \bar{{\cal J}}^\mu
      (\varrho,u)
        :=\lambda^2 M^+{\cal H}(\varrho,u) M^{\circ*} 
           + \lambda^2 M^\circ{\cal H}(\varrho,u) M^{+*} 
           + M^+{\cal H}(\varrho,u) + {\cal H}(\varrho,u) M^{+*}, 
\nonumber \\ & & \hspace*{-1em}
    {\cal H}(\varrho,u):={\rm e}^{\mu_1\lambda^2 K(u)/2}
                 \varrho\hspace{0.1em}
                    {\rm e}^{\mu_1\lambda^2 K(u)/2}. 
\nonumber
\end{eqnarray}
Eq.\ \eqref{zakai-SME} is a simple recursion for 
a $2\times 2$ matrix and is thus easily implementable 
on a digital computer. Note that the operators 
$\bar{{\cal L}}^\mu$ and $\bar{{\cal J}}^\mu$ 
reduce when $\mu_1=0$ to 
$\bar{{\cal L}}^0=\bar{{\cal L}}$ and 
$\bar{{\cal J}}^0=\bar{{\cal J}}$, 
where $\bar{{\cal L}}$ and $\bar{{\cal J}}$ are given in Eqs. 
(\ref{def-of-Lbar}) and (\ref{def-of-Jbar}). 
This implies that the solution of Eq.\ \eqref{zakai-SME} 
converges  to that of Eq.\ \eqref{Q-zakai-infor-state} 
when $\mu_1$ goes to zero.

%%%%%%%%%%%%%%%%%%%%%%%%%%%%%%%%%%%%%%%%%%%%%%%%%%%%%%%%%%%%%%%%%%%
%%%%%%%%%%%%%%%%%%%%%%%%%%%%%%%%%%%%%%%%%%%%%%%%%%%%%%%%%%%%%%%%%%%

\subsection{Calculating the risk-sensitive estimator}

We will now represent the cost function 
$F$ of Eq.\ \eqref{costfunction} 
in terms of $\varrho_l^\mu$ only. To this end, 
we define a new state ${\mathbb Q}^l$ on 
$\mathcal{M}\otimes\mathcal{W}_N$ by 
\begin{equation*}
\label{new-state}
    {\mathbb Q}^l(X):={\mathbb P}\big[U(l)XU(l)^*\big], 
\end{equation*}
for all $X \in \mathcal{M}\otimes \mathcal{W}_N$. 
Since $\mathcal{Y}_l$ is a commutative $*$-subalgebra 
of $\mathcal{M}\otimes \mathcal{W}_N$, we can 
apply Theorem \ref{Spectral theorem} 
to $(\mathcal{Y}_l,\mathbb{Q}^l)$. 
That is, there exists a classical probability space 
$(\Omega_l,\mathcal{F}_l,\mathbf{Q}^l)$ and a $*$-isomorphism 
$\iota:\ \mathcal{Y}_l \to \ell^\infty(\mathcal{F}_l)$ such 
that $\mathbb{Q}^l(A) = \mathbf{E}_{\mathbf{Q}^l}[\iota(A)]$ for all 
$A \in \mathcal{Y}_l$. We now have the following theorem.

\begin{theorem}
\label{cost representation by info state}
The cost function in Eq.\ \eqref{costfunction} 
can be written as  
\begin{equation}
\label{rs-minimizer}
    F({\mathfrak u}_1,\ldots,{\mathfrak u}_l)
     ={\bf E}_{{\bf Q}^l}\Big[
           \Tr\big(\varrho^\mu_l
                 {\rm e}^{\mu_2|X_{{\rm e}}-u_l|^2} \big) \Big], 
\end{equation}
where $u_l=\iota({\mathfrak u}_l)$ is a function of the measurement data 
$\Delta y_1,\ldots,\Delta y_l$, and $\varrho^\mu_l$ is the risk-sensitive 
information density matrix that satisfies Eq.\ \eqref{zakai-SME}. 
\end{theorem}

\begin{proof}
We define ${\check {\mathfrak u}}_l=j_l^{-1}({\mathfrak u}_l)\in{\cal C}_l$ 
as before. 
Since $|j_l(X_{{\rm e}})-{\mathfrak u}_l|^2
=j_l(|X_{{\rm e}}-{\check {\mathfrak u}}_l|^2)$, we find
\begin{eqnarray}
& & \hspace*{-1em}
    F({\mathfrak u}_1,\ldots,{\mathfrak u}_l)
      ={\mathbb P}\Big[R(l)^*
          {\rm e}^{\mu_2 j_l(|X_{{\rm e}}
                              -{\check {\mathfrak u}}_l|^2)}R(l)\Big]
      ={\mathbb P}\Big[R(l)^*U(l)^*
          {\rm e}^{\mu_2|X_{{\rm e}}
                              -{\check {\mathfrak u}}_l|^2}U(l) R(l)\Big]
\nonumber \\ & & \hspace*{4.8em}
      ={\mathbb P}\Big[U^\mu(l)^*
          {\rm e}^{\mu_2|X_{{\rm e}}
                              -{\check {\mathfrak u}}_l|^2}U^\mu(l)\Big], 
\nonumber
\end{eqnarray}
where $U^\mu(l)$ is defined by Eq.\ \eqref{def-of-Umu}. 
Using Eq.\ \eqref{V-tildeV} in Lemma \ref{U-V connection} 
and the tower property 
of the conditional expectation, we find 
\begin{eqnarray}
& & \hspace*{-1em}
    F({\mathfrak u}_1,\ldots,{\mathfrak u}_l)
      ={\mathbb P}\Big[V^\mu(l)^*
         {\rm e}^{\mu_2|X_{{\rm e}}-{\check {\mathfrak u}}_l|^2}V^\mu(l)\Big]
      ={\mathbb P}\Big[ 
         {\mathbb P}\Big(V^\mu(l)^*
             {\rm e}^{\mu_2|X_{{\rm e}}-{\check {\mathfrak u}}_l|^2}V^\mu(l)
                      \condbigbar{\cal C}_l\Big)\Big]
\nonumber \\ & & \hspace*{4.85em}
    ={\mathbb Q}^l\Big[ U(l)^*
         {\mathbb P}\Big(V^\mu(l)^*
               {\rm e}^{\mu_2|X_{{\rm e}}-{\check {\mathfrak u}}_l|^2}V^\mu(l)
                      \condbigbar{\cal C}_l\Big) U(l)\Big]
    ={\mathbb Q}^l\Big[ \sigma_l^\mu
        \Big({\rm e}^{\mu_2|X_{{\rm e}}
             -{\check {\mathfrak u}}_l|^2}\Big)\Big], 
\nonumber
\end{eqnarray}
where we have used the definition 
of $\sigma_l^\mu$ in Eq.\ \eqref{def-of-sigma}. 
Note that the above conditional expectation is well defined due to 
$X_{{\rm e}}-{\check {\mathfrak u}}_l\in{\cal C}_l'$ and 
$V^\mu(l)\in{\cal C}_l'$.  
Let $u_l$ be given by 
$u_l=\iota(j_l({\check {\mathfrak u}}_l))=\iota({\mathfrak u}_l)$.
It now follows from Lemma \ref{info-composition} that
\[
     \iota\Big( \sigma^\mu_l(
             {\rm e}^{\mu_2|X_{{\rm e}}-{\check {\mathfrak u}}_l|^2}) \Big)
      =\Tr\Big[\varrho^\mu_l{\rm e}^{\mu_2|X_{{\rm e}}-u_l|^2}\Big]. 
\]
Consequently, the cost function can be written as 
\[
    F({\mathfrak u}_1,\ldots,{\mathfrak u}_l)
     ={\bf E}_{{\bf Q}^l}
          \Big[\iota\Big(\sigma^\mu_l(
              {\rm e}^{\mu_2|X_{{\rm e}}
                          -{\check {\mathfrak u}}_l|^2})\Big)\Big]
     ={\bf E}_{{\bf Q}^l}
          \Big[\Tr\Big(\varrho^\mu_l{\rm e}^{\mu_2|X_{{\rm e}}
                          -u_l|^2}\Big)\Big]. 
\]
This completes the proof. 
\end{proof}

As a result of Theorem \ref{cost representation by info state}, 
our estimation problem 
is now cast as a classical optimal control problem. 
The resulting problem can be solved systematically 
by dynamic programming. We will only provide a brief 
summary of this. Consider the following {\it optimal 
expected cost-to-go} $f_l(\varrho)$ at time $l$, given that 
$\varrho^\mu_l = \varrho$ 
\[
    f_l(\varrho)
      =\min_{u_l,\ldots,u_N}
        {\bf E}_{{\bf Q}^N}\Big[
           \Tr\big[\varrho^\mu_N
                 {\rm e}^{\mu_2|X_{{\rm e}}-u_N|^2} \big] \condbigbar 
                      \varrho^\mu_l=\varrho \Big],~~
    f_{N+1}(\varrho)=\Tr(\varrho). 
\]
This leads to the following dynamic programming equation; 
denoting Eq.\ \eqref{zakai-SME} simply as 
$\varrho^\mu_l=\Gamma(\varrho^\mu_{l-1},u_{l-1},\Delta y_l)$, we have 
\[
    f_{l-1}(\varrho)
     =\min_{u_{l-1}}{\bf E}_{{\bf Q}^N}
        \Big[f_l(\Gamma(\varrho,u_{l-1},\Delta y_l))\Big]
     =\min_{u_{l-1}}\half
        \Big[f_l(\Gamma(\varrho,u_{l-1},\lambda))
            +f_l(\Gamma(\varrho,u_{l-1},-\lambda))\Big]. 
\]
Note here that 
${\rm Prob}_{{\bf Q}^N}(\Delta y_l=\lambda)={\mathbb Q}^N[U(l)^*P_l^+U(l)]
={\mathbb P}(P_l^+)=1/2$. 
We can run the above algorithm efficiently in a digital computer and obtain 
the optimal sequence $u_l^{{\rm opt}}~(l=1,\ldots,N)$, which yields 
$\widehat{X_{{\rm e}}}^\mu(l)=\iota^{-1}(u_l^{{\rm opt}})$ through a 
verification theorem (e.g., see \cite{james3}). Theorem \ref{robustness-1} 
below will lead to a robustness result for the risk-sensitive 
estimator.

\begin{remark}
\label{suboptimal estimator}

Running the dynamic programming recursion on a digital 
computer is very costly computationally. Therefore 
we define a {\it suboptimal risk-sensitive estimator} 
by
\begin{equation}
\label{suboptimal-rs-filter}
    \widehat{X_{{\rm e}}}^{\mu, {\rm sub}}(l)
      :=\mathop{\mathrm{argmin}}_{
          {\mathfrak u}_l\in{\cal Y}_l}
             F\left(\widehat{X_{{\rm e}}}^{\mu, {\rm sub}}(1),
                      \ldots,\widehat{X_{{\rm e}}}^{\mu, {\rm sub}}(l-1), 
               {\mathfrak u}_l\right). 
\end{equation}
That is, $\widehat{X_{{\rm e}}}^{\mu, {\rm sub}}(l)$ is to be calculated 
based on the assumption that we have already performed the above 
minimization procedure up to time $l-1$ and obtained the suboptimal 
risk-sensitive estimators 
$\widehat{X_{{\rm e}}}^{\mu, {\rm sub}}(i)\in{\cal Y}_i~(i=1,\ldots,l-1)$. 
As shown in \cite{boel} (Theorems 2.2 and 4.2), the minimizer of the 
trace function inside the expectation in Eq.\ \eqref{rs-minimizer}, 
$u_l^{{\rm min}}$, leads to the suboptimal risk-sensitive estimator 
$\widehat{X_{{\rm e}}}^{\mu, {\rm sub}}(l)=\iota^{-1}(u_l^{{\rm min}})$. 
Hence our algorithm in this case is represented simply as follows: 
\begin{equation}
\label{rs-suboptimal-minimizer}
    \iota\left(\widehat{X_{{\rm e}}}^{\mu, {\rm sub}}(l)\right)
     =\mathop{\mathrm{argmin}}_{u_l\in{\bf R}}
         \Tr\left[\varrho^\mu_l
            {\rm e}^{\mu_2|X_{{\rm e}}-u_l|^2} \right],~~~
    \varrho^\mu_l
      =\Gamma\left(\varrho^\mu_{l-1},
          \iota\left(\widehat{X_{{\rm e}}}^{\mu, {\rm sub}}(l-1)\right),
                                                      \Delta y_l\right), 
\end{equation}
which is of the same structure as the 
classical algorithm presented 
in Section I-A. In Theorem \ref{robustness-2} 
we will derive a bound for the conditional estimation error.  
The suboptimal risk-sensitive estimator minimizes 
this error bound. This provides a sound theoretical 
foundation for the suboptimal risk-sensitive estimator. 
Since the algorithm (\ref{rs-suboptimal-minimizer}) 
is computationally much cheaper than 
the dynamic programming equation, we will consistently use 
the suboptimal estimator in the example part, Section VII.
\end{remark}

%%%%%%%%%%%%%%%%%%%%%%%%%%%%%%%%%%%%%%%%%%%%%%%%%%%%%%%%%%%%%%%%%%%
%%%%%%%%%%%%%%%%% Q uncertain systems and robustness %%%%%%%%%%%%%%
%%%%%%%%%%%%%%%%%%%%%%%%%%%%%%%%%%%%%%%%%%%%%%%%%%%%%%%%%%%%%%%%%%%

\section{Quantum uncertain systems and robustness of the risk-sensitive 
filter}

In realistic situations, we often have to 
deal with a system that includes 
some model uncertainty. 
From the classical case, we expect 
that the risk-sensitive estimator  
has an enhanced robustness 
property against such uncertainty. 
In this section, we first describe a 
class of uncertain quantum systems 
for which the uncertainty is quantified 
by the quantum relative entropy. 
We will then show robustness properties 
of the estimator.

%%%%%%%%%%%%%%%%%%%%%%%%%%%%%%%%%%%%%%%%%%%%%%%%%%%%%%%%%%%%%%%%%%%
%%%%%%%%%%%%%%%%%%%%%%%%%%%%%%%%%%%%%%%%%%%%%%%%%%%%%%%%%%%%%%%%%%%

\subsection{Quantum uncertain systems}

Uncertainty can enter the system in many ways. 
It could for instance be the case that the 
state $\psi$ is unknown to us. The uncertainty 
then enters the system density matrix $\rho$ 
through the relation ${\mathbb P}(X)=\psi\otimes\phi^{\otimes N}(X)
=\Tr[X(\rho\otimes(\Phi\Phi^*)^{\otimes N})]$. We assume that the 
field state is known and fixed to the vacuum $\phi$.
This, however, is not the only way uncertainty 
can enter our model. We will also allow for uncertainty 
in the coefficients of the dynamics, 
i.e., the difference equation \eqref{discreteQsde}. 
We can push this uncertainty into the 
initial state, as described below.

Let $(\Omega,\mathcal{F},\mathbf{P})$ be a classical 
probability space. Let $p$ be an element of 
$L^\infty(\Omega,\mathcal{F},\mathbf{P})$, i.e.\ $p$ 
is a random variable on $(\Omega,\mathcal{F},\mathbf{P})$. 
Let $\mathfrak{p}$ be the operator on 
$L^2(\Omega,\mathcal{F},\mathbf{P})$ given 
by pointwise multiplication with $p$, i.e., 
  \begin{equation*}
  (\mathfrak{p}f)(\omega) = p(\omega)f(\omega),~~
  f \in L^2(\Omega,\mathcal{F},\mathbf{P}),\ \omega \in \Omega. 
  \end{equation*}
We denote the commutative $*$-algebra of 
all such multiplication operators with functions 
in $L^\infty(\Omega,\mathcal{F},\mathbf{P})$
by $\mathcal{P}$. On $\mathcal{P}$ we can 
define a state $\tau$ as integration 
with respect to the measure $\mathbf{P}$, 
i.e., $\tau(\mathfrak{p}) = 
\int_\Omega p(\omega)\mathbf{P}(d\omega)$.
For simplicity we will take the operator 
$\mathfrak{p}$ to be self-adjoint, i.e.\  
it is a multiplication with a real-valued 
function. 
Next, let $M^i,~(i=\pm,+,-,\circ)$ be $\mathcal{M}$-valued 
functions on $\mathbb{C}$, i.e.\ 
$M^i: \mathbb{C} \to \mathcal{M}$, such that 
the matrix $M_l$ in Eq.\ \eqref{M-evolution} is unitary. 
Then, using the compositions of $M^i$ and 
$\mathfrak{p} \in \mathcal{P}$, we can define 
the following difference equation 
  \begin{equation}
  \label{uncertain-discreteQsde}
    \Delta U(l)
      =\big[
       M^{\pm}({\mathfrak p})\Delta\Lambda(l) 
       + M^{+}({\mathfrak p})\Delta A(l)^* 
       + M^{-}({\mathfrak p})\Delta A(l) 
       + M^{\circ}({\mathfrak p})\Delta t(l) \big] U(l-1),~~
    U(0)=I
  \end{equation}
on the extended quantum probability space 
\[
    ({\cal P}\otimes{\cal M}\otimes{\cal W}_N, {\mathbb P})
    =({\cal P}\otimes{\cal M}\otimes{\cal M}^{\otimes N}, 
       \tau\otimes\psi\otimes\phi^{\otimes N}). 
\]
We now assume that the state 
$\tau\otimes\psi$ is unknown to us. 
This means that Eq.\ \eqref{uncertain-discreteQsde} 
is equivalent to the difference equation \eqref{discreteQsde} 
such that its coefficients include uncertain parameter $p$. 
That is, the uncertainty in the model 
has been pushed completely into the 
state $\tau\otimes\psi$.

Now, let 
$\rho^{{\rm true}}
=\rho_{{\rm p}}^{{\rm true}}\otimes\rho_{{\rm s}}^{{\rm true}}$ 
be the {\it true density matrix} corresponding to the unknown state 
$\tau\otimes\psi$. 
Then, the true filter is initialized to $\varrho_0=\rho^{{\rm true}}$. 
However, as $\rho^{{\rm true}}$ is unknown, we fix a {\it nominal 
density matrix} 
$\rho^{{\rm nom}}=\rho_{{\rm p}}^{{\rm nom}}
\otimes\rho_{{\rm s}}^{{\rm nom}}$, which in general differs from 
$\rho^{{\rm true}}$, and construct the nominal 
filter that starts from $\varrho_0=\rho^{{\rm nom}}$. 
The nominal estimator of $j_l(X)$ is then given by 
\[
    \pi^{{\rm nom}}_l(X)
    =\frac{\sigma^{{\rm nom}}_l(X)}{\sigma^{{\rm nom}}_l(I)},~~
    \sigma^{{\rm nom}}_l(X)
      =\iota^{-1}\Big(\Tr(\varrho^{{\rm nom}}_lX)\Big),~~
    \varrho_0=\rho^{{\rm nom}}. 
\]

\begin{example}
\label{uncertain example}
If $p$ is a discrete random variable that 
takes the values $p_i~(i=1,\ldots,m)$ with 
unknown probability $r_i$, then the corresponding 
multiplication operator is 
${\mathfrak p}={\rm diag}\{p_1,\ldots,p_m\}$ $\in{\cal P}$, 
where the commutative $*$-algebra ${\cal P}$ 
is the set of $m\times m$ diagonal 
matrices, and the true density matrix is 
$\rho^{{\rm true}}_{{\rm p}}={\rm diag}\{r_1,\ldots,r_m\}$. 
To design a nominal filter, we choose a nominal density matrix 
of the form $\rho^{{\rm nom}}_{{\rm p}}={\rm diag}\{r'_1,\ldots,r'_m\}$. 
In general, $r_i\neq r_i'$. 
It is easily seen that the quantum 
relative entropy between the above two 
distributions is equal to the classical one: 
\[
     R(\rho^{{\rm true}}_{{\rm p}}\|\rho^{{\rm nom}}_{{\rm p}})
      =\sum_{i=1}^m r_i\log\frac{r_i}{r_i'}
      =R^c(\{r_i\}\|\{r_i'\}). 
\]
Note that an important example for a true density matrix is 
$\rho^{{\rm true}}_{{\rm p}}={\rm diag}\{1,0,\ldots,0\}$; 
that is, $p$ is not a random variable but an unknown deterministic 
system parameter $p=p_1$. 
If we have no information about $p$ at all, it 
is natural to take a uniform distribution
$\rho^{{\rm nom}}_{{\rm p}}={\rm diag}\{1/m, \ldots, 1/m\}$ 
as the nominal distribution.
\end{example}

%%%%%%%%%%%%%%%%%%%%%%%%%%%%%%%%%%%%%%%%%%%%%%%%%%%%%%%%%%%%%%%%%%%
%%%%%%%%%%%%%%%%%%%%%%%%%%%%%%%%%%%%%%%%%%%%%%%%%%%%%%%%%%%%%%%%%%%

\subsection{Robustness properties of the risk-sensitive and suboptimal 
risk-sensitive estimators}

The nominal estimator $\pi^{{\rm nom}}_l(X_{{\rm e}})$ differs 
from the true one $\pi^{{\rm true}}_l(X_{{\rm e}})$. 
Hence, $\pi^{{\rm nom}}_l(X_{{\rm e}})$ is no longer the optimal 
estimator in the sense of the mean square error and thus can possibly 
take a large estimation error. 
However, as shown below, if one uses the 
nominal risk-sensitive estimator (given by
Eq.\ \eqref{def-of-rs-estimator}), the estimation 
error is guaranteed to be within a certain bound. 
This implies that the risk-sensitive estimator 
does have a robustness property against unknown 
perturbation of the system state and the system 
parameters.

The quantum relative entropy (\ref{QrelativeEntropy}) 
will be used to express the robustness property. 
We here assume that the unknown true density matrix 
$\rho^{{\rm true}}
=\rho_{{\rm p}}^{{\rm true}}\otimes\rho_{{\rm s}}^{{\rm true}}$ is 
within a certain distance from a known nominal density matrix 
$\rho^{{\rm nom}}
=\rho_{{\rm p}}^{{\rm nom}}\otimes\rho_{{\rm s}}^{{\rm nom}}$: 
\begin{equation}
\label{assumption-entropy}
    R(\rho^{{\rm true}}\|\rho^{{\rm nom}})<+\infty. 
\end{equation}
The following theorem will lead to a robustness 
property of the nominal risk-sensitive estimator 
defined in Eq.\ \eqref{def-of-rs-estimator}.

\begin{theorem}
\label{robustness-1}
Let ${\mathfrak u}_l~(l=1,\ldots,N)$ 
be an element of ${\mathcal Y}_l$. 
Then, we have the following inequality 
\begin{eqnarray}
\label{filter-robustness}
& & \hspace*{-1em}
    {\mathbb P}_{{\rm true}}\Big[
      \log\Big( R(N)^*{\rm e}^{\mu_2|j_N(X_{{\rm e}})-{\mathfrak u}_N|^2} 
      R(N)  \Big) \Big]
\nonumber \\ & & \hspace*{2em}
    \leq
      \log{\mathbb P}_{{\rm nom}}\Big[
        R(N)^*{\rm e}^{\mu_2|j_N(X_{{\rm e}})-{\mathfrak u}_N|^2} R(N) \Big]
        +R(\rho^{{\rm true}}\|\rho^{{\rm nom}}), 
\end{eqnarray}
where $R(N)$ is defined by Eq.\ \eqref{def-of-R}, and 
${\mathbb P}_{{\rm true}}(X)
=\Tr[X(\rho^{{\rm true}}\otimes(\Phi\Phi^*)^{\otimes N})]$ and 
${\mathbb P}_{{\rm nom}}(X)
=\Tr[X(\rho^{{\rm nom}}\otimes(\Phi\Phi^*)^{\otimes N})]$. 
\end{theorem}

\begin{proof}
Setting $\rho=\rho^{{\rm true}}\otimes(\Phi\Phi^*)^{\otimes N}$ and 
$\rho'=\rho^{{\rm nom}}\otimes(\Phi\Phi^*)^{\otimes N}$ in Eq.\ 
\eqref{duality-ineq}, we have 
\[
    {\mathbb P}_{{\rm true}}(Z)
    \leq \log{\mathbb P}_{{\rm nom}}({\rm e}^Z)
       +R(\rho^{{\rm true}}\|\rho^{{\rm nom}}),~~
    \forall Z\in{\cal P}\otimes{\cal M}\otimes{\cal W}_N, 
\]
where we have used the following additivity property: 
\begin{eqnarray}
& & \hspace*{-1em}
    R(\rho^{{\rm true}}\otimes(\Phi\Phi^*)^{\otimes N}
                 \|\rho^{{\rm nom}}\otimes(\Phi\Phi^*)^{\otimes N})
    =R(\rho^{{\rm true}}\|\rho^{{\rm nom}})
        +R((\Phi\Phi^*)^{\otimes N}\|(\Phi\Phi^*)^{\otimes N})
\nonumber \\ & & \hspace*{15.9em}
    =R(\rho^{{\rm true}}\|\rho^{{\rm nom}}). 
\nonumber
\end{eqnarray}
Therefore, taking
$Z=\log[R(N)^*{\rm e}^{\mu_2|j_N(X_{{\rm e}})-{\mathfrak u}_N|^2} R(N)]$ 
yields the theorem. 
\end{proof}

Eq.\ \eqref{filter-robustness} is a quantum version of the classical 
robustness result \eqref{classical-robustness}, because the left hand 
side of Eq.\ \eqref{filter-robustness} can be expanded up to second order 
in the estimation error as 
\begin{eqnarray}
& & \hspace*{-1em}
    {\mathbb P}_{{\rm true}}\Big[
        \mu_1\sum_{l=1}^{N-1}|j_l(X_{{\rm e}})-{\mathfrak u}_l|^2
                       +\mu_2|j_N(X_{{\rm e}})-{\mathfrak u}_N|^2 \Big]
                  +O(|j_l(X_{{\rm e}})-{\mathfrak u}_l|^4)
\nonumber \\ & & \hspace*{0em}
    \leq
      \log{\mathbb P}_{{\rm nom}}\Big[
        R(N)^*{\rm e}^{\mu_2|j_N(X_{{\rm e}})-{\mathfrak u}_N|^2} R(N) \Big]
     \mbox{}
        +R(\rho^{{\rm true}}\|\rho^{{\rm nom}}). 
\nonumber
\end{eqnarray}
That is, as in the classical case, the nominal 
risk-sensitive estimator 
${\mathfrak u}_l=\widehat{X_{{\rm e}}}^{\mu, {\rm nom}}(l)$, 
defined by Eq.\ \eqref{def-of-rs-estimator}, 
does have a robustness property, because it minimizes 
the upper bound of the estimation error under the 
unknown true state ${\mathbb P}_{{\rm true}}$.

We remark that the relative entropy in 
Eq.\ \eqref{filter-robustness} can be written 
as 
\[
  R(\rho^{{\rm true}}\|\rho^{{\rm nom}})
   =R(\rho^{{\rm true}}_{{\rm p}}\|\rho^{{\rm nom}}_{{\rm p}})
   +R(\rho^{{\rm true}}_{{\rm s}}\|\rho^{{\rm nom}}_{{\rm s}}). 
\]
The first term is a classical relative entropy 
as shown in Example \ref{uncertain example}. 
Thus, if there is no uncertainty in the quantum state, 
the estimation error bound is written in terms of 
classical quantities only.

We now change our focus to the suboptimal risk-sensitive 
estimator $\widehat{X_{{\rm e}}}^{\mu, {\rm sub}}(l)$ 
defined in Remark \ref{suboptimal estimator}.
The following Theorem shows that the conditional estimation error 
at time $l$ also has an upper bound. This will lead 
to a robustness property for the nominal suboptimal 
risk-sensitive estimator of Remark \ref{suboptimal estimator}.

\begin{theorem}
\label{robustness-2}
Let ${\mathfrak u}_l~(l=1,\ldots,N)$ 
be an element of ${\mathcal Y}_l$. 
Then, we have the following inequality 
\begin{equation}
\label{filter-robustness-conditional}
    \iota\Big( {\mathbb P}_{{\rm true}}
      \big[|j_l(X_{{\rm e}})-{\mathfrak u}_l|^2\big|{\cal Y}_l\big] \Big)
        \leq \frac{1}{\mu_2}
              \log\Tr\big[\rho^{\mu,{\rm nom}}_l
                            {\rm e}^{\mu_2|X_{{\rm e}}-u_l|^2} \big]
        +\frac{1}{\mu_2}R(\rho_l^{{\rm true}}\|\rho_l^{\mu,{\rm nom}}), 
\end{equation}
where 
$\rho_l^{{\rm true}}=\varrho_l^{{\rm true}}/\Tr[\varrho_l^{{\rm true}}]$ 
and 
$\rho_l^{\mu,{\rm nom}}
=\varrho_l^{\mu,{\rm nom}}/\Tr[\varrho_l^{\mu,{\rm nom}}]$ 
are the conditional density matrices corresponding to the true filter 
and the nominal risk-sensitive filter, and $u_l=\iota({\mathfrak u}_l)$. 
\end{theorem}

\begin{proof}
Using the definition of the optimal estimator $\pi_l(X_{{\rm e}})$, 
the left-hand side in inequality 
(\ref{filter-robustness-conditional}) can be 
rewritten as  
\[
    \iota\Big({\mathbb P}_{{\rm true}}
      \big[j_l(|X_{{\rm e}}-\check{{\mathfrak u}}_l|^2)
        \big|{\cal Y}_l\big] \Big)
    =\iota\Big(\pi_l^{{\rm true}}
           \big(|X_{{\rm e}}-\check{{\mathfrak u}}_l|^2\big)\Big)
    =\Tr\Big[ \rho_l^{{\rm true}}|X_{{\rm e}}-u_l|^2 \Big], 
\]
where the last equality follows directly from 
Lemma \ref{info-composition} with 
$u_l=\iota(j_l({\check {\mathfrak u}}_l))
=\iota({\mathfrak u}_l)$. 
Then, from Eq.\ \eqref{duality-ineq} we have the assertion. 
\end{proof}

The first term of the right-hand 
side in Eq.\ \eqref{filter-robustness-conditional} is 
minimized when choosing the nominal suboptimal 
risk-sensitive estimator 
$u_l=\iota\left(\widehat{X_{{\rm e}}}^{\mu, {\rm sub}}(l)\right)$ 
given by Eq.\ \eqref{rs-suboptimal-minimizer}.
Theorem \ref{robustness-2} therefore shows a robustness property 
of the suboptimal risk-sensitive estimator defined 
in Remark \ref{suboptimal estimator}.

%%%%%%%%%%%%%%%%%%%%%%%%%%%%%%%%%%%%%%%%%%%%%%%%%%%%%%%%%%%%%%%%%%%
%%%%%%%%%%%%%%%%%%%%%%%%%%% Simulation %%%%%%%%%%%%%%%%%%%%%%%%%%%%
%%%%%%%%%%%%%%%%%%%%%%%%%%%%%%%%%%%%%%%%%%%%%%%%%%%%%%%%%%%%%%%%%%%

\section{Examples}\label{sec examples}

In this section, we study 
two examples in detail. 
The first example is a two-level
atom that is coupled to the field 
via a {\it dispersive interaction}. 
This coupling can be obtained by putting 
the atom in a cavity that has a resonance 
frequency far detuned from the transition 
frequency of the two-level atom.
The second example deals with a two-level 
atom that decays to the ground state 
due to {\it spontaneous emission} into 
its environment. 
We consider the situation 
where the quantum state of the two-level atom 
and a physical parameter are unknown to us. 
In particular, we employ the nominal suboptimal 
risk-sensitive estimator given by  
Eq.\ \eqref{rs-suboptimal-minimizer}.
We compare this estimator with both the 
true risk-neutral and nominal 
risk-neutral estimators.

%%%%%%%%%%%%%%%%%%%%%%%%%%%%%%%%%%%%%%%%%%%%%%%%%%%%%%%%%%%%%%%%%%%
%%%%%%%%%%%%%%%%%%%%%%%%%%%%%%%%%%%%%%%%%%%%%%%%%%%%%%%%%%%%%%%%%%%

\subsection{Dispersive interaction model}

The interaction Hamiltonian (\ref{hamiltonian}) in 
case of a dispersive interaction with the field, 
is given by the following system matrices: 
\begin{equation}
\label{nominal-Ls}
    L_1=0,~~
    L_2=i\sqrt{g}\sigma_z,~~
    L_3=0, 
\end{equation}
where $\sigma_z={\rm diag}\{1,-1\}$ 
and $g>0$ represents the interaction 
strength. From Eqs.\ \eqref{hamiltonian-unitary} 
and \eqref{M-evolution}, we see that the matrices 
$M^i~(i=\pm,+,-,\circ)$ are given by 
\[
    M^\pm(g)=0,~~
    M^+(g)=\frac{\sin(g\lambda)}{\lambda}\sigma_z,~~
    M^-(g)=-\frac{\sin(g\lambda)}{\lambda}\sigma_z,~~
    M^\circ(g)=\frac{\cos(g\lambda)-1}{\lambda^2}I. 
\]
We assume that $g$ is a classical random variable 
that takes the values $g_i$ with unknown probabilities 
${\rm Prob}(g_i)=r_i$. 
As seen in Section VI-A, $g$ can be regarded as an 
observable ${\mathfrak g}={\rm diag}\{g_1,\ldots,g_m\}\in{\cal P}$, 
where ${\cal P}$ is a commutative $*$-algebra given by the 
set of $m\times m$ diagonal matrices. 
The corresponding unknown true density matrix is 
$\rho^{{\rm true}}_{{\rm p}}={\rm diag}\{r_1,\ldots,r_m\}$. 
In particular, we now study a toy model in which $g$ 
can take $20$ discrete values, 
$g_i=0.4+0.03i~(i=1,\ldots,20)$. Moreover, we choose
$\rho^{{\rm true}}_{{\rm p}}$ 
to be given by 
\begin{equation}
\label{true-density-1}
     \rho^{{\rm true}}_{{\rm p}}
      ={\rm diag}\{0, 0.01, 0.04, 0.1, 0.7, 0.1, 0.04, 0.01, 0, 0, 
                   0, 0,    0,    0,   0,   0,   0,    0,    0, 0\}, 
\end{equation}
which is illustrated in Fig. 1 (a1). 
For instance, $g$ takes 
$g_3=0.49$ with probability
${\rm Prob}(g_3)=0.04$. 
Furthermore, we assume 
that the true density matrix is given by 
\begin{equation}
\label{true-density-2}
    \rho^{{\rm true}}_{{\rm s}}
     =\left[ \begin{array}{cc}
               0.5 & 0.5 \\
               0.5 & 0.5 \\
             \end{array} \right]. 
\end{equation}
Again, note that 
$\rho^{{\rm true}}
=\rho^{{\rm true}}_{{\rm p}}\otimes\rho^{{\rm true}}_{{\rm s}}$ 
is unknown to us.

Now, let us consider estimating 
the system observable $X_{{\rm e}}=\sigma_z$. 
To design a nominal filter, we use the 
following nominal density matrix 
in ${\cal P}\otimes{\cal M}$: 
\begin{equation}
\label{nominal-density-1}
   \rho^{{\rm nom}}
   =\rho^{{\rm nom}}_{{\rm p}}\otimes\rho^{{\rm nom}}_{{\rm s}},~~~
   \rho^{{\rm nom}}_{{\rm p}}
   ={\rm diag}\Big\{\frac{1}{20},\ldots,\frac{1}{20}\Big\},~~~
   \rho^{{\rm nom}}_{{\rm s}}
   =\left[ \begin{array}{cc}
          0.5 & 0.25 \\
          0.25 & 0.5 \\
    \end{array} \right]. 
\end{equation}
$\rho^{{\rm nom}}_{{\rm p}}$ is depicted in Fig.~1 (a2). 
The nominal risk-neutral estimator 
$\pi_l^{{\rm nom}}(\sigma_z)$ and the  
nominal risk-sensitive one 
$\widehat{\sigma_z}^{\mu, {\rm sub}}(l)$ 
are then calculated from Eq.\ \eqref{Q-zakai-infor-state} 
with $\varrho_0=\rho^{{\rm nom}}$ and 
Eq.\ \eqref{rs-suboptimal-minimizer} 
with $\varrho^\mu_0=\rho^{{\rm nom}}$, respectively. 
The risk-sensitive parameters are chosen to 
be $(\mu_1, \mu_2)=(0.1, 0.182)$. 
Note that the filter equations include the composition $M^i({\mathfrak g})$ 
and are driven by the true output data $\Delta y_l$. 
We compare those two nominal estimators with 
the ideal true risk-neutral estimator 
$\pi_l^{{\rm true}}(\sigma_z)$, which is calculated from Eq.\ 
\eqref{Q-zakai-infor-state} with $\varrho_0=\rho^{{\rm true}}$. 
To do this, we use the averaged total estimation errors 
\begin{equation}
\label{averaged total estimation error}
    \Delta^{{\rm rn}}=
    \frac{1}{N}\sum_{l=1}^N\big|
       \iota(\pi_l^{{\rm true}}(\sigma_z))
           -\iota(\pi_l^{{\rm nom}}(\sigma_z))\big|,~~~
    \Delta^{{\rm rs}}=
    \frac{1}{N}\sum_{l=1}^N\left|
       \iota(\pi_l^{{\rm true}}(\sigma_z))
           -\iota\left(\widehat{\sigma_z}^{\mu, {\rm sub}}(l)\right)\right|. 
\end{equation}
The histogram for these values are depicted 
in Fig. 1 (b) for $200$ sample paths with 
$\lambda^2=0.001$ and $N=2000$. 
Overall, $\Delta^{{\rm rs}}$ is smaller 
than $\Delta^{{\rm rn}}$, showing the better 
performance of the risk-sensitive estimator 
over the risk-neutral one. 
Figs.~1 (c) and (d) illustrate an example of 
sample paths of the estimators; 
in Fig.~1 (c) the solid line shows 
$\iota(\widehat{\sigma_z}^{\mu, {\rm sub}}(l))$, 
while in Fig.~1 (d) the solid line is 
$\iota(\pi_l^{{\rm nom}}(\sigma_z))$. 
In both figures, the thick dotted line is 
$\iota(\pi_l^{{\rm true}}(\sigma_z))$. 
In Fig.~1 (c), both estimators are quite close 
to each other in spite of the difference in 
their initial states. 
On the other hand, as depicted in Fig.~1 (d), the 
nominal risk-neutral estimator fails in the 
estimation, although it finally converges to the 
true value $-1$. 
As a summary, the risk-sensitive estimator 
outperforms the nominal risk-neutral estimator 
in the presence of uncertainty.

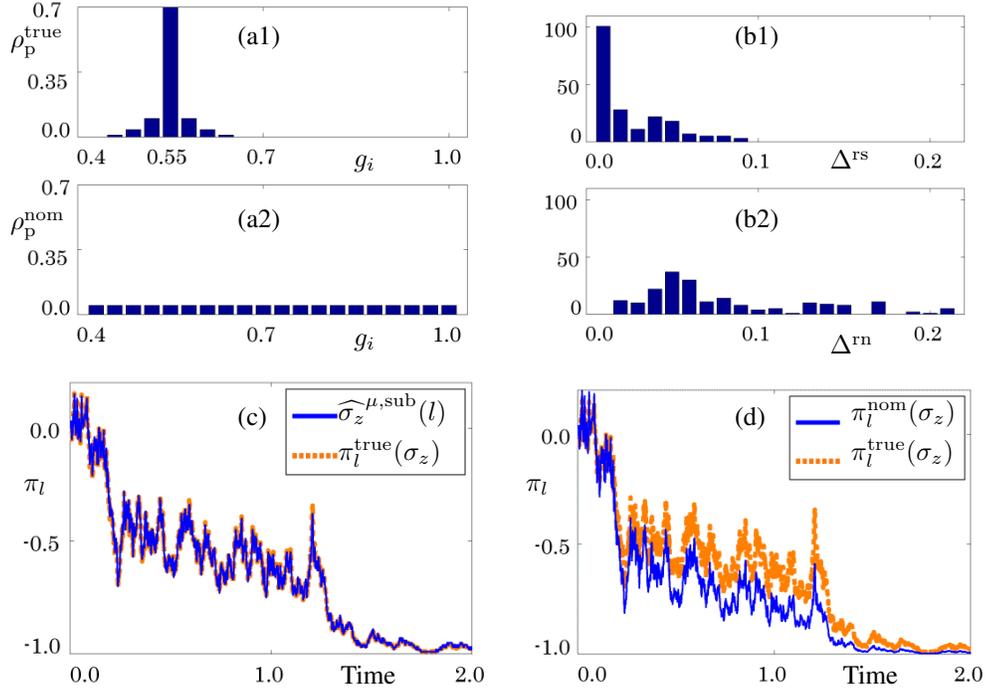
\begin{figure}[t]
\centering
\begin{picture}(400,270)

%\put(12,129)
%{\includegraphics[width=2.4in]{DisProbBar}}
%\put(204,130)
%{\includegraphics[width=2.35in]{DisBar}}
%\put(0,0)
%{\includegraphics[width=2.65in]{dispersiveR}}
%\put(200,3)
%{\includegraphics[width=2.5in]{dispersiveN}}

\put(11,129)
{\includegraphics[width=2.63in]{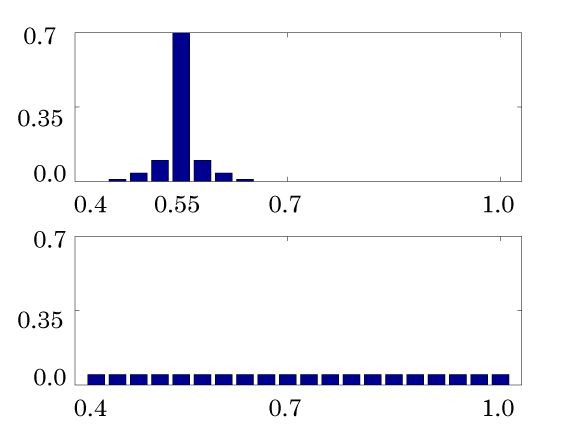}}
\put(204,130)
{\includegraphics[width=2.55in]{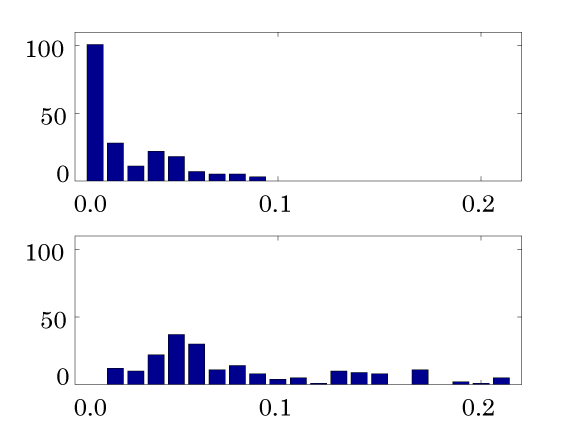}}
\put(1,0)
{\includegraphics[width=2.8in]{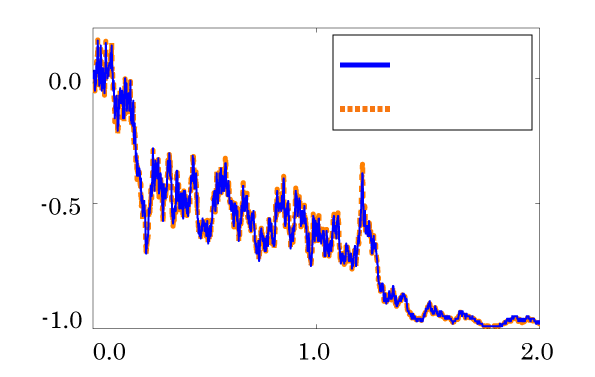}}
\put(200,3)
{\includegraphics[width=2.65in]{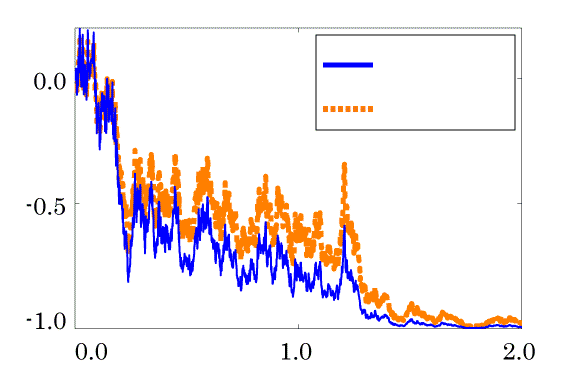}}

%\put(100,80){$\pi_l^{{\rm true}}(\sigma_z)$}
%\put(50,25){$\pi_l^{\mu,{\rm nom}}(\sigma_z)$}
\put(134,90){$\pi_l^{{\rm true}}(\sigma_z)$}
\put(134,105){$\widehat{\sigma_z}^{\mu, {\rm sub}}(l)$}
\put(328,90){$\pi_l^{{\rm true}}(\sigma_z)$}
\put(328,105){$\pi_l^{{\rm nom}}(\sigma_z)$}

\put(15,78){$\pi_l$}
\put(205,78){$\pi_l$}

\put(10,177){$\rho^{{\rm nom}}_{{\rm p}}$}
\put(10,246){$\rho^{{\rm true}}_{{\rm p}}$}
\put(140,201){$g_i$}
\put(140,133){$g_i$}

\put(135,5){Time}
\put(325,5){Time}
\put(320,200){$\Delta^{{\rm rs}}$}
\put(320,130){$\Delta^{{\rm rn}}$}

\put(96,247){(a1)}
\put(96,178){(a2)}
\put(284,247){(b1)}
\put(284,178){(b2)}
\put(96,103){(c)}
\put(284,103){(d)}

\end{picture}

\caption{
For the dispersive interaction model of the atom, 
(a) the true and nominal parameter distributions, 
(b) the histogram of the averaged total estimation errors, 
(c) sample paths of the nominal risk-sensitive estimator (solid line) 
and the true risk-neutral one (thick dotted line), and 
(d) sample paths of the nominal risk-neutral estimator (solid line) and 
the true risk-neutral one (thick dotted line). 
For the figures (c) and (d), the notation ($\iota$) is omitted. 
} 
\end{figure}

\begin{remark}
The performance of the nominal estimator 
depends on the magnitude of uncertainty. 
For example, if there is no uncertainty 
in the nominal distribution, the nominal 
risk-neutral estimator coincides with the true optimal estimator 
and clearly works better than the risk-sensitive one. 
However, under the existence of some uncertainty, 
the risk-neutral estimator is no longer optimal 
and will be inferior to the risk-sensitive one, as seen 
in Fig.~1. 
To make a more quantitative observation, we consider the following nominal 
distribution characterized by one parameter $\beta\in[0,1]$ that represents 
the uncertainty magnitude: 
\begin{eqnarray}
& & \hspace*{-1em}
    (\rho_{{\rm p}}^{{\rm nom},\beta})_{1,1}
    =(\rho_{{\rm p}}^{{\rm nom},\beta})_{9,9}
    =\ldots
    =(\rho_{{\rm p}}^{{\rm nom},\beta})_{20,20}
    =0.05\beta, 
\nonumber \\ & & \hspace*{-1em}
    (\rho_{{\rm p}}^{{\rm nom},\beta})_{2,2}
    =(\rho_{{\rm p}}^{{\rm nom},\beta})_{8,8}
    =0.04\beta+0.01,~~~
    (\rho_{{\rm p}}^{{\rm nom},\beta})_{3,3}
    =(\rho_{{\rm p}}^{{\rm nom},\beta})_{7,7}
    =0.01\beta+0.04, 
\nonumber \\ & & \hspace*{-1em}
    (\rho_{{\rm p}}^{{\rm nom},\beta})_{4,4}
    =(\rho_{{\rm p}}^{{\rm nom},\beta})_{6,6}
    =-0.05\beta+0.1,~~~
    (\rho_{{\rm p}}^{{\rm nom},\beta})_{5,5}
    =-0.65\beta+0.7, 
\nonumber \\ & & \hspace*{-1em}
   \rho_{{\rm s}}^{{\rm nom},\beta}
   =\left[ \begin{array}{cc}
          0.5 & 0.5-0.25\beta \\
          0.5-0.25\beta & 0.5 \\
    \end{array} \right]. 
\nonumber
\end{eqnarray}
When $\beta=0$, the nominal distribution is equal to the true one; 
$\rho^{{\rm nom},0}_{{\rm p}}\otimes\rho^{{\rm nom},0}_{{\rm s}}
=\rho^{{\rm true}}_{{\rm p}}\otimes\rho^{{\rm true}}_{{\rm s}}$. 
Hence $\beta=0$ implies there is no uncertainty. 
On the other hand when $\beta=1$, the nominal distribution is the one 
given in Eq.\ \eqref{nominal-density-1}. 
We consider the nominal risk-neutral estimator and the risk-sensitive 
one with $\mu_1=0.01, \mu_2=0.05$. 
Note that these two estimators are close to 
each other due to the small risk-sensitive 
parameter. To evaluate their performances, 
we calculate the averaged total estimation 
errors \eqref{averaged total estimation error} 
and compare them. 
In Fig.~2 (a1), the horizontal axis shows the uncertainty 
magnitude $\beta$, while the vertical axis shows the average of 
$\Delta^{{\rm rn}}$ and $\Delta^{{\rm rs}}$ over $100$ sample paths, 
which are denoted by $\bar{\Delta}^{{\rm rn}}$ and 
$\bar{\Delta}^{{\rm rs}}$, respectively. 
Fig.~2 (a2) shows examples of the nominal parameter distribution 
$\rho^{{\rm nom},\beta}_{{\rm p}}$. 
The risk-sensitive estimator clearly shows a better 
performance than the risk-neutral one, except in 
the case of a small $\beta$. 
\end{remark}

\begin{remark}
The robustness property of the risk-sensitive 
filter is based on the fact that the estimation error 
is upper bounded, as presented in 
Theorems \ref{robustness-1} and \ref{robustness-2}. 
Fig.~2 (b) illustrates sample paths of the conditional 
estimation error and its upper bound given in Theorem 
\ref{robustness-2}:
\begin{equation}
\label{Th62revisit}
    \varepsilon_l:=
    \iota\Big( {\mathbb P}_{{\rm true}}
      \big[|j_l(X_{\rm e})-{\mathfrak u}_l|^2\big|{\cal Y}_l\big] \Big),~~
    \varepsilon_l':=
    \frac{1}{\mu_2}
              \log\Tr\big[\rho^{\mu,{\rm nom}}_l
                            {\rm e}^{\mu_2|X_{\rm e}-u_l|^2} \big]
        +\frac{1}{\mu_2}R(\rho_l^{{\rm true}}\|\rho_l^{\mu,{\rm nom}}). 
\end{equation}
From this, we see that the bound 
is much larger than the actual estimation error. This 
is very similar to the classical case where one also 
often finds a very conservative upper bound.  
\end{remark}

\begin{remark}
As in the classical case, there is no 
theoretical procedure to determine 
the best risk-sensitive parameters $(\mu_1, \mu_2)$. 
We here only maintain that a non-zero 
$\mu_1$, the weighting parameter 
of the running estimation error cost, 
is actually helpful in obtaining a 
high-quality risk-sensitive filter. 
To show this fact, we apply a nominal 
risk-sensitive filter with $\mu_1=0$, 
initialized to Eq.\ \eqref{nominal-density-1}, 
to the same uncertain system as discussed above. 
For this filter, $\mu_2=0.281$ appears to 
be the best parameter. Fig.\ 2 (c) illustrates 
the mean values over $200$ sample paths of the 
conditional estimation error $\varepsilon_l$ in 
Eq.\ \eqref{Th62revisit}. 
The upper dotted line and lower 
solid one corresponds to the risk-sensitive 
estimation with $(\mu_1,\mu_2)=(0.0, 0.281)$ and 
$(\mu_1,\mu_2)=(0.1, 0.182)$, respectively. 
This shows that a non-zero $\mu_1$ does improve 
the performance of the estimator. 
\end{remark}

\begin{figure}[t]
\centering
\begin{picture}(400,255)

%\put(7,130)
%{\includegraphics[width=2.5in]{uncertainmag}}
%\put(204,130)
%{\includegraphics[width=2.35in]{DisProbGs}}
%\put(14,0)
%{\includegraphics[width=2.4in]{Derror}}
%\put(204,0)
%{\includegraphics[width=2.4in]{mucomp}}

\put(0,130)
{\includegraphics[width=2.72in]{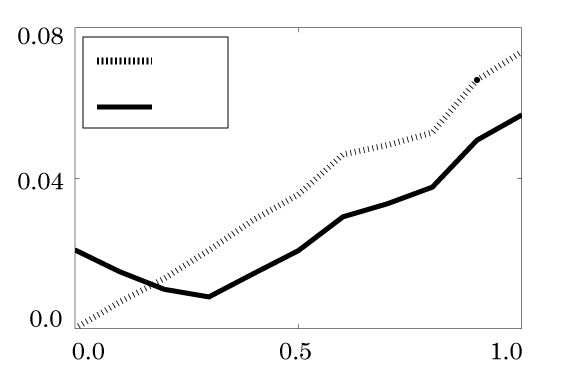}}
\put(197,130)
{\includegraphics[width=2.65in]{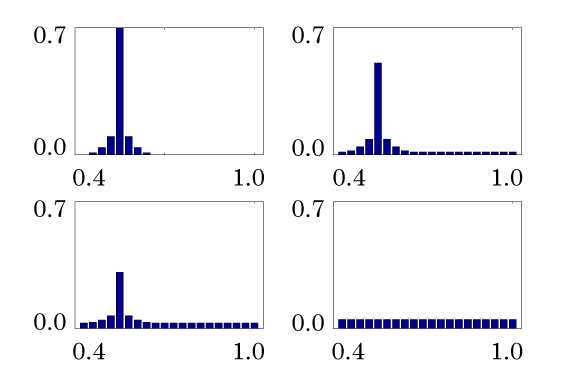}}
\put(0,0)
{\includegraphics[width=2.7in]{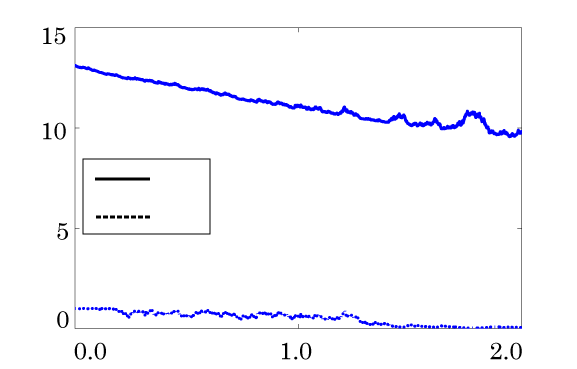}}
\put(197,0)
{\includegraphics[width=2.65in]{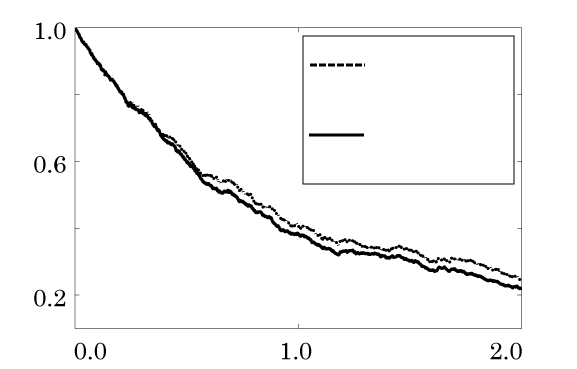}}

\put(12,215){$\bar{\Delta}$}
\put(92,233){(a1)}
\put(140,130){$\beta$}
\put(60,231){$\bar{\Delta}^{{\rm rn}}$}
\put(60,215){$\bar{\Delta}^{{\rm rs}}$}

\put(196,220){$\rho^{{\rm nom}}_{{\rm p}}$}
\put(196,160){$\rho^{{\rm nom}}_{{\rm p}}$}
\put(248,132){$g_i$}
\put(338,132){$g_i$}

\put(257,235){(a21)}
\put(345,235){(a22)}
\put(257,175){(a23)}
\put(345,175){(a24)}
\put(243,210){$\beta=0.0$}
\put(333,210){$\beta=0.3$}
\put(243,150){$\beta=0.6$}
\put(333,150){$\beta=1.0$}

\put(92,102){(b)}
\put(15,94){$\varepsilon$}
\put(60,49){$\varepsilon_l$}
\put(60,62){$\varepsilon_l'$}
\put(135,0){Time}

\put(279,100){(c)}
\put(206,90){$\varepsilon_l$}
\put(325,0){Time}
\put(318,66){$\mu_1=0.1$}
\put(318,76){$\mu_2=0.182$}
\put(318,90){$\mu_1=0.0$}
\put(318,100){$\mu_2=0.281$}

\end{picture}

\caption{
For the dispersive interaction model of the atom, 
(a1) the averaged total estimation errors of the nominal risk-sensitive 
filter (solid line) and the nominal risk-neutral one (dotted line), 
(a2) examples of the nominal parameter distributions, 
(b) the conditional error (dotted line) and the guaranteed error bound 
(solid line) in Theorem \ref{robustness-2}, and 
(c) the averaged conditional estimation errors of the nominal 
risk-sensitive filters with $(\mu_1, \mu_2)=(0.0, 0.281)$ (upper dotted line) 
and $(\mu_1, \mu_2)=(0.1, 0.182)$ (lower solid line). 
} 
\end{figure}
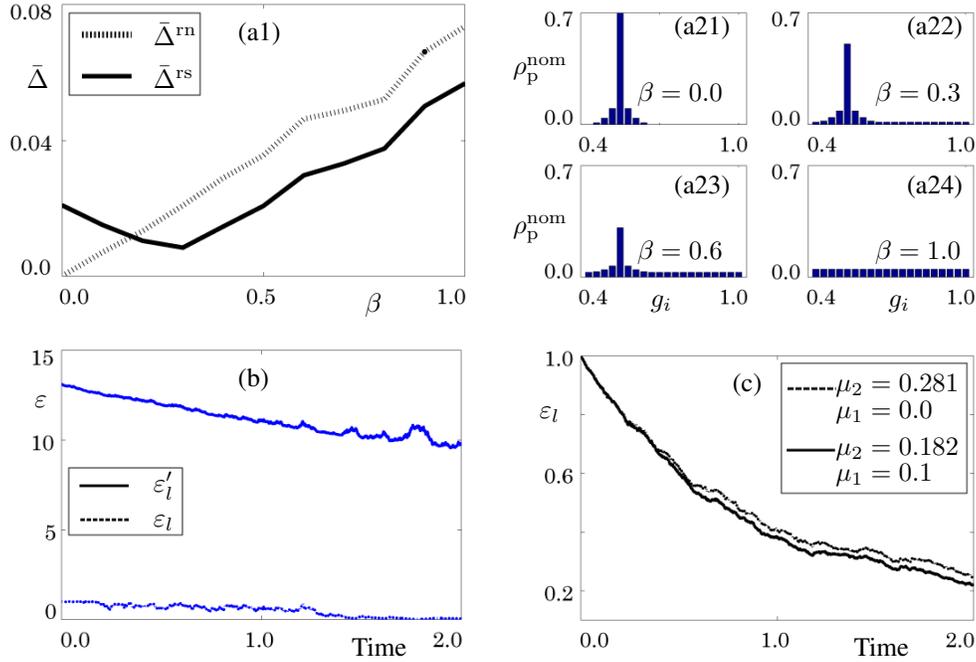

%%%%%%%%%%%%%%%%%%%%%%%%%%%%%%%%%%%%%%%%%%%%%%%%%%%%%%%%%%%%%%%%%%%
%%%%%%%%%%%%%%%%%%%%%%%%%%%%%%%%%%%%%%%%%%%%%%%%%%%%%%%%%%%%%%%%%%%

\subsection{Spontaneous emission model}

In the case of spontaneous decay, 
the interaction Hamiltonian 
(\ref{hamiltonian}) is given by 
\begin{equation}
\label{nominal-Ls-Emis}
    L_1=0,~~
    L_2=i\sqrt{e}\sigma_-,~~
    L_3=0, 
\end{equation}
where $\sigma_-$ is defined in Eq.\ \eqref{creation-annihilation} and $e>0$ 
represents the emission rate. 
The matrices $M^i~(i=\pm,+,-,\circ)$ are determined from Eqs. 
(\ref{hamiltonian-unitary}) and (\ref{M-evolution}) and read 
\begin{eqnarray}
& & \hspace*{-1em}
    M^\pm(e)=(1-\cos(e\lambda))\sigma_z,~~
    M^+(e)=\frac{\sin(e\lambda)}{\lambda}\sigma_-,
\nonumber \\ & & \hspace*{-1em}
    M^-(e)=-\frac{\sin(e\lambda)}{\lambda}\sigma_+,~~
    M^\circ(e)=\frac{\cos(e\lambda)-1}{\lambda^2}\sigma_+\sigma_-. 
\nonumber
\end{eqnarray}
Similar to the dispersive interaction case, 
we here assume that $e$ behaves as a classical 
discrete random variable with an unknown probability 
distribution; 
$e$ is then replaced by an observable 
${\mathfrak e}={\rm diag}\{e_1,\ldots,e_m\}\in{\cal P}$. 
In particular, we assume 
$e_i=0.2+0.04i~(i=1,\ldots,20)$ with the true density matrix 
\[
     \rho^{{\rm true}}_{{\rm p}}
      ={\rm diag}\{0, 0, 0, 0, 0, 0, 0, 0, 0, 0, 
                   0, 0, 0, 0, 0.01, 0.04, 0.9, 0.04, 0.01, 0\}, 
\]
which is illustrated in Fig.~3 (a). 
The system true density matrix $\rho_{{\rm s}}^{{\rm true}}$ 
is given by Eq.\ \eqref{true-density-2}. 
For a nominal density matrix, we take $\rho_{{\rm p}}^{{\rm nom}}$ 
in Eq.\ \eqref{nominal-density-1} and assume that 
$\rho_{{\rm s}}^{{\rm nom}}=\rho_{{\rm s}}^{{\rm true}}$. 
In the above setting, we consider estimating 
$X_{{\rm e}}=\sigma_y:=i(\sigma_--\sigma_+)$, 
investigate the performance of 
the nominal risk-sensitive filter, and 
compare it with the nominal risk-neutral one. 
The risk-sensitive parameters are chosen as 
$(\mu_1,\mu_2)=(0.15, 0.25)$. 
Fig.~3 (b) shows the histogram for the averaged 
error $\Delta^{{\rm rs}}$ and $\Delta^{{\rm rn}}$ 
for $200$ sample paths with $\lambda^2=0.001$ and 
$N=5000$. 
Fig.~3 (c) shows the sample paths of 
$\iota(\pi_l^{{\rm true}}(\sigma_y))$ and 
$\iota(\widehat{\sigma_y}^{\mu, {\rm sub}}(l))$, while 
in Fig.~3 (d) $\iota(\pi_l^{{\rm true}}(\sigma_y))$ 
and $\iota(\pi^{\rm nom}_l(\sigma_y))$ are shown. 
These figures clearly show that the nominal 
risk-sensitive estimator is superior to the nominal 
risk-neutral estimator.

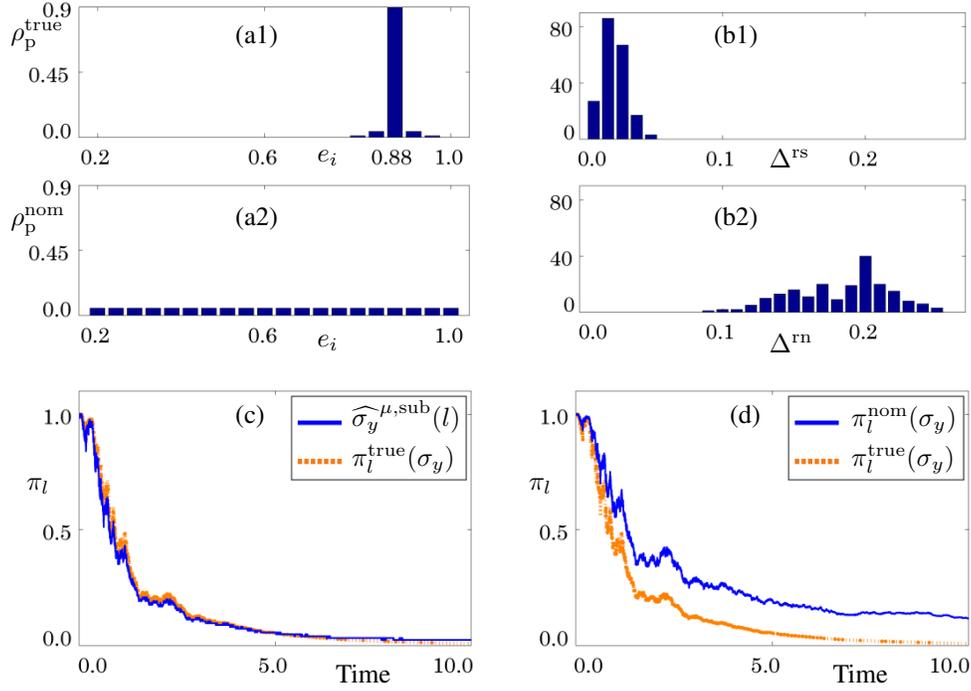
\begin{figure}[t]
\centering
\begin{picture}(400,275)

%\put(10,130)
%{\includegraphics[width=2.43in]{SpoProbBar}}
%\put(205,130)
%{\includegraphics[width=2.35in]{SponBar}}
%\put(14,3)
%{\includegraphics[width=2.5in]{SpoRiskLong}}
%%\put(198,3)
%\put(203,3)
%{\includegraphics[width=2.5in]{SpoNeutLong}}

\put(10,128)
{\includegraphics[width=2.64in]{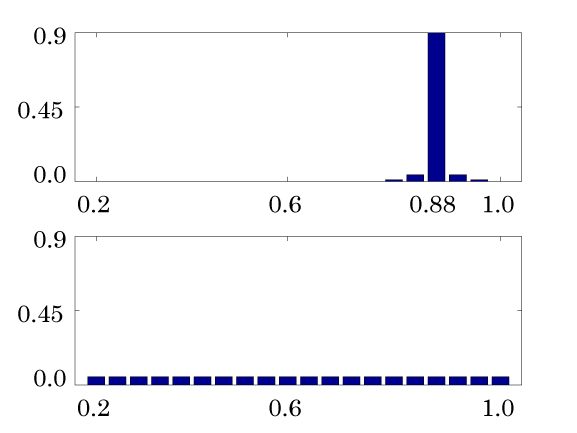}}
\put(200,130)
{\includegraphics[width=2.6in]{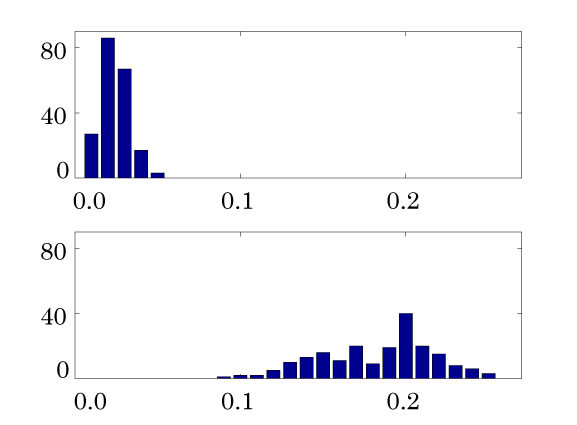}}
\put(10,6)
{\includegraphics[width=2.65in]{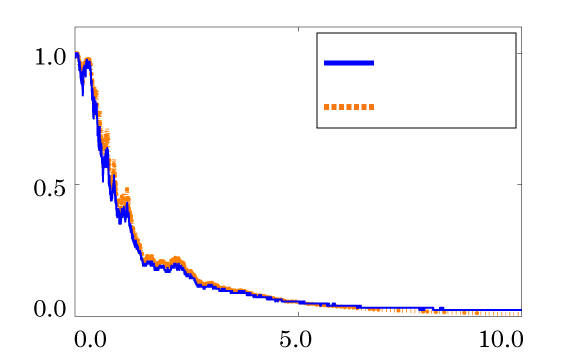}}
\put(198,6)
{\includegraphics[width=2.65in]{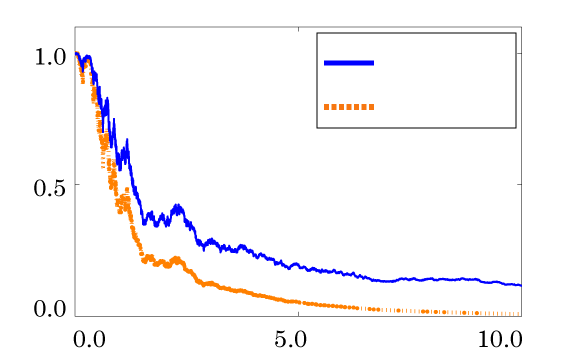}}

\put(9,178){$\rho^{{\rm nom}}_{{\rm p}}$}
\put(9,248){$\rho^{{\rm true}}_{{\rm p}}$}
\put(125,202){$e_i$}
\put(125,132){$e_i$}

\put(132,5){Time}
\put(320,5){Time}

\put(296,200){$\Delta^{{\rm rs}}$}
\put(296,130){$\Delta^{{\rm rn}}$}

\put(138,87){$\pi_l^{{\rm true}}(\sigma_y)$}
\put(138,102){$\widehat{\sigma_y}^{\mu, {\rm sub}}(l)$}
\put(327,87){$\pi_l^{{\rm true}}(\sigma_y)$}
\put(327,102){$\pi_l^{{\rm nom}}(\sigma_y)$}

\put(15,78){$\pi_l$}
\put(205,78){$\pi_l$}

\put(94,247){(a1)}
\put(94,178){(a2)}
\put(275,247){(b1)}
\put(275,178){(b2)}
\put(94,103){(c)}
\put(281,103){(d)}

\end{picture}

\caption{
For the spontaneous emission model of the atom, 
(a) the true and nominal parameter distributions, 
(b) the histogram of the averaged total estimation errors, 
(c) sample paths of the nominal risk-sensitive estimator (solid line) 
and the true risk-neutral one (thick dotted line), and 
(d) sample paths of the nominal risk-neutral estimator (solid line) and 
the true risk-neutral one (thick dotted line). 
For the figures (c) and (d), the notation ($\iota$) is omitted. 
} 
\end{figure}

\begin{remark} 
While in this paper we have considered the estimation 
problem over the finite-time horizon, let us here look 
at the asymptotic behaviour as $l\rightarrow\infty$ of 
the following quantity
\[
    \delta_l
       =|\pi^{{\rm true}}_l(X_{\rm e})-\pi^{{\rm nom}}_l(X_{\rm e})|, 
\]
where $\pi^{{\rm true}}_l(X_{\rm e})$ and 
$\pi^{{\rm nom}}_l(X_{\rm e})$ correspond to the standard 
risk-neutral estimator for the true and nominal initial 
states, respectively. 
If $\lim_{l\rightarrow\infty}\delta_l=0$ for all 
observables $X_{\rm e}$, then we say the filter is {\it stable}. 
Recently, Van Handel \cite{ramon2} has provided the following  
characterization for filter stability 
in continuous time. For all 
$X_{\rm e}$ included in the {\it observable space}
\[
    {\cal O}={\rm span}
              \{ {\cal L}^{c_1}{\cal J}^{d_1}{\cal L}^{c_2}
                \cdots{\cal L}^{c_k}{\cal J}^{d_k}(I)~:~k,c_i,d_i\geq 0 \},
\]
we have $\delta_l\rightarrow 0$. 
Here, ${\cal L}$ and ${\cal J}$ are the continuous 
time analogues of the quantities defined in Eq.\ \eqref{lindbladian}. 
Therefore, the filter is stable if 
${\rm dim}{\cal O}={\rm dim}{\cal A}$.

In our examples the observable spaces are 
given by
\[
    {\cal O}^{{\rm dis}}={\rm span}\{I,\ \sigma_z\},~~~
    {\cal O}^{{\rm spon}}={\rm span}\{I,\ \sigma_x,\ \sigma_z\}. 
\]
Therefore, for a dispersive interaction where we 
estimate $\sigma_z\in{\cal O}^{{\rm dis}}$, it is 
guaranteed by Van Handel's theorem that 
$\pi^{{\rm nom}}_l(\sigma_z)$ with any initial state 
converges to the true estimator. 
On the other hand, in the spontaneous decay case, due to 
$\sigma_y\notin{\cal O}^{{\rm spon}}$, we cannot 
expect that $\delta_l\rightarrow 0$. 
This could be the reason why the 
increase in performance by the nominal 
risk-sensitive estimator over the risk-neutral 
one is more pronounced in Fig.~3 than 
in Fig.~1. We must note here that 
in simulations we do see that, 
with the settings used in Fig.~3, 
$\pi^{{\rm nom}}_l(\sigma_y)$ 
eventually converges to the true value $0$.  
However, this convergence is very slow.

\end{remark}

%%%%%%%%%%%%%%%%%%%%%%%%%%%%%%%%%%%%%%%%%%%%%%%%%%%%%%%%%%%%%%%%%%%
%%%%%%%%%%%%%%%%%%%%%%%%%%% Conclusion %%%%%%%%%%%%%%%%%%%%%%%%%%%%
%%%%%%%%%%%%%%%%%%%%%%%%%%%%%%%%%%%%%%%%%%%%%%%%%%%%%%%%%%%%%%%%%%%

%\section{Conclusion}

%\appendices





%\section{}


\section*{Acknowledgment}

We thank I. R. Petersen and V. A. Ugrinovskii for 
insightful discussion and for bringing the quantum 
duality relation to our attention. 
We also thank M. R. James for helpful comments on 
simulation results. 
NY acknowledges support by the JSPS Grant-in-Aid No. 
06693. 
LB acknowledges support by the ARO under
Grant No. W911NF-06-1-0378.

\end{document}